\documentclass[]{aa}  
\usepackage[colorlinks]{}
\usepackage{graphicx}
\usepackage[]{natbib}
\usepackage[switch]{lineno}
\bibpunct{(}{)}{;}{a}{}{,} % to follow the A&A style
\usepackage{txfonts}
\usepackage{amstext}

\begin{document}

   \title{Solar flares observed by Rosetta at comet 67P}

   %\subtitle{}https://www.overleaf.com/18743208nnjphbgsmfqs

   \author{N. J. T. Edberg \inst{1}
          \and
          F. L. Johansson  \inst{1,2}
          \and
          A. I. Eriksson \inst{1}
          \and
          D. J. Andrews \inst{1}
          \and
          R. Hajra\thanks{Now at the National Atmospheric Research Laboratory, Gadanki 517112, India} \inst{3}
          \and
          P. Henri \inst{3}
          \and
          C. Simon Wedlund \inst{4}
          \and
          M. Alho \inst{5}
          \and
          E. Thiemann \inst{6}
}

\institute{Swedish Institute of Space Physics, Uppsala, Sweden, \email{ne@irfu.se}
         \and
         Department of Physics and Astronomy, Uppsala University, Uppsala, Sweden
         \and
        Laboratoire de Physique et Chimie de l’Environnement et de l’Espace (LPC2E), CNRS, Orléans, France 
        \and
                Department of Physics, University of Oslo, Box 1048 Blindern, 0316 Oslo, Norway
        \and Department of Radio Science and Engineering, School of Electrical Engineering, Aalto University, Aalto, Finland
        \and
        Laboratory for Atmospheric and Space Physics, University of Colorado, 3665 Discovery Drive Boulder, CO 80303, USA
}

   \date{Received}

%\linenumbers
% \abstract{}{}{}{}{} 
% 5 {} token are mandatory
 
  \abstract
   {The Rosetta spacecraft made continuous measurements of the coma of comet 67P/Churyumov-Gerasimenko (67P) for more than two years. The plasma in the coma appeared very dynamic, and many factors control its variability.}
   {We wish to identify the effects of solar flares on the comet plasma and also their effect on the measurements by the Langmuir Probe Instrument (LAP).}
   {To identify the effects of flares, we proceeded from an existing flare catalog of Earth-directed solar flares, from which a new list was created that only included Rosetta-directed flares.  We also used measurements of flares at Mars when at similar longitudes as Rosetta. The flare irradiance spectral model (FISM v.1) and its Mars equivalent (FISM-M) produce an extreme-ultraviolet (EUV) irradiance (10-120 nm) of the flares at 1 min resolution. LAP data and density measurements obtained with the Mutual Impedence Probe (MIP) from the time of arrival of the flares at Rosetta were examined to determine the flare effects.}
   {From the vantage point of Earth, 1504 flares directed toward Rosetta occurred during the mission. In only  24 of these, that is, 1.6\%, was the increase in EUV irradiance large enough  to cause an observable effect in LAP data. Twenty-four Mars-directed flares were also observed in Rosetta data. The effect of the flares was to increase the photoelectron current by typically 1-5 nA. We find little evidence that the solar flares increase the plasma density, at least not above the background variability.}
   {Solar flares have a small effect on the photoelectron current of the LAP instrument, and they are not significant in comparison to other factors that control the plasma density in the coma. The photoelectron current can only be used for flare detection during periods of calm plasma conditions.}


   \maketitle
%
%________________________________________________________________

\section{Introduction}
Solar flares can have significant effects on planetary atmospheres because the X-ray increase can heat up the thermosphere and the extreme-ultraviolet (EUV) increase can boost the ionization rate in the ionospheres \citep[e.g.,][]{tsurutani2009, mendillo2006, thiemann2015, lee2018}. For low- and intermediate-activity comets with atmospheres of low neutral density compared to planets, flares are expected to mainly affect processes that are governed by the EUV radiation, such as ionization of neutrals and photoelectrons emitted from illuminated surfaces. The study presented in this paper is therefore aimed at determining both the efficiency of solar flares as a transient solar forcing mechanism on cometary ionospheres and their effect on the photoelectron emission from the Langmuir Probe Instrument (LAP). This study is complementary to previous studies of other types of transient solar forcing, for instance, solar wind pressures pulses such as coronal mass ejections (CMEs) or corotating interaction regions (CIR), which have been shown to have large effects on the plasma environment of comet 67P \citep{edberg2016, edberg2016b, hajra2018, noonan2018, goetz2018}. The results will also complement the understanding of the variability of a cometary plasma environment, which is affected by a number of short-time scale processes (minutes to hours) that include, for example, cometary outbursts \citep{grun2016, hajra2017}, the changing amount of cold plasma \citep{eriksson2016, engelhardt2018}, plasma waves \citep{volwerk2016, andre2017}, and transient structures in the magnetic field, such as current sheets and the diamagnetic cavity \citep{volwerk2017, goetz2016}.

Solar flares originate in the solar corona, typically close to sunspots. They are believed to be created as a result of magnetic reconnection, which accelerates and heats plasma particles. As the accelerated particles follow magnetic field lines down to their footprints in the photosphere, they  eventually lead to severe heating through collisions with the denser plasma farther down in the solar atmosphere, and the plasma wells up. This plasma then fills the coronal loops and cools down, whereas the particles loose parts of their energy by emitting radiation at EUV as well as X-ray wavelengths \citep[e.g.,][]{thiemann2018}. The X-ray radiation typically comes first, and as the plasma cools further, the EUV radiation follows, with a time delay of up to tens of minutes \citep{thiemann2017}. The radiation from a flare is isotropic (because of the free-free emission process) and spreads hemispherically; it hits any celestial body in sight, that is, a body located in the same hemisphere. Because the coronal structure is dense, some limb darkening of flares can occur when they originate from regions that lie far out on the limb of the Sun, such that they have to pass through some of the solar atmosphere. This mainly affects those parts of the EUV wavelengths ($\sim$25-120 nm) where emission lines and continua are optically thick \citep{qian2009, thiemann2018}.

The Rosetta spacecraft arrived at comet 67P in 2014. At slow walking pace ($\sim$m/s), it investigated the near nucleus neutral and plasma environment for two years, following the comet through its perihelion passage and outward in the solar system again. The trajectory of Rosetta around the nucleus varied throughout the mission such that it spanned large portions of the cometary latitudes and longitudes as well as solar zenith angles (although never in eclipse behind the comet).  As 67P is an intermediately active comet with a gas production rate of $\sim 10^{25} - 10^{29}$ s$^{-1}$ \citep{hansen2016}, the neutral density at Rosetta's location never became optically thick to EUV or X-rays, for instance. Throughout the mission, the neutral and plasma density at Rosetta varied by many orders of magnitude, spanning some $10^6-10^9$ cm$^{-3}$ and $10^1-10^4$ cm$^{-3}$, respectively \citep[e.g.,][]{heritier2018}. 

Because a solar flare is a frequent phenomenon on the Sun, several can be observed each day if the active region on the Sun is favorably directed \citep{veronig2002}. There is also a solar cycle dependency such that they are more frequent at solar maximum and less so during solar minimum.  The Rosetta spacecraft arrived at the comet in 2014 shortly after solar maximum, and it stayed in orbit until October 2016 during the declining phase of solar cycle 24. Unfortunately, the number of sunspots during this cycle was unusually low and relatively few intense flares occurred. The intensity of  solar flares varies from case to case, and the emitted power spans several orders of magnitude from the weakest to the most intense flares. The classification scheme commonly used is based on the peak flux of X-rays in the wavelength interval 0.1-0.8 nm, as measured by the Geostationary Operational Environmental Satellite (GOES) spacecraft. The different categories B, C, M, and X indicate if the magnitude of the peak flux is greater than $10^{-7}$, $10^{-6}$, $10^{-5}$ , and $10^{-4}$ W/m${^2}$, respectively. However, there is not a simple relation between the intensity in X-ray emission and the emission in EUV, meaning that this classification is not necessarily appropriate for the EUV part of the flare, that is, to evaluate the efficiency of the flare in terms of ionization. This has implications for characterizing a planetary body's response to solar flares.  For example, \citet{le2012} showed that flare-induced neutral density enhancements in Earth's upper atmosphere were approximately twice better correlated to peak 26-34 nm EUV irradiance than to peak 0.1-0.8 nm soft X-ray irradiance.

In the following Section 2 we describe the relevant instruments on Rosetta, in Section 3 we describe how solar flares are measured at Earth and Mars, and in Section 4 we present a number of observations of flares, as well as non-detections, at Rosetta. We conclude the paper with a discussion followed by our conclusions.

\section{Rosetta instruments}
There is no dedicated solar flare monitor on Rosetta, but an estimate of the integrated EUV flux reaching the spacecraft is possible to obtain from the LAP instrument by measuring the amount of photoelectrons coming off it when illuminated by sunlight  \citep{eriksson2007, johansson2017}. The LAP instrument consists of two TiN coated spherical Langmuir probes of radius 2.5 cm, mounted on stubs on booms that extend approximately 2.2 and 1.6 m away from the spacecraft, respectively. In the most commonly used mode, a bias voltage is applied to the probe and depending on the sign of the probe potential, free electrons or ions in the ambient space are attracted or repelled. The bias potential can either be swept in voltage steps, from a maximum of -32 to +32 V at a cadence of $\sim$ 1-3 min, or be set to a fixed value whereby the current can be collected at a higher time resolution; we used 32 s here. 

When the probe is sunlit, photoelectrons are emitted from the TiN surface. These photoelectrons add to the total current measured by the probe, and this contribution is proportional to the integrated EUV flux. To measure the maximum saturated photoelectron current, $I_{ph0}$,  the probe needs to be negatively charged with respect to the plasma. Otherwise the emitted photoelectrons would be attracted back to the probe by the positive potential, resulting in a decrease of the measured current. The absolute value of $I_{ph0}$ can be estimated, for instance, by determining the change in measured current as the probe moves in and out of shadow during spacecraft manoeuvres. It can also be estimated by studying the derivative of the collected current at negative potential with respect to the bias potential for a collection of individual sweeps as introduced by \citet{johansson2017}.  These methods are not useful for studying solar flares because their time-resolution does not extend to $\text{}$minutes. Instead, we need to use the direct current measurement from the probe and look for changes that occur over the interval of the flare length. This also requires that there are no other changes in the plasma density that would change the collected current. This criterion is unfortunately rarely fulfilled in the very dynamic plasma environment of comet 67P. Still, we use the current collected at a negative bias potential here to identify changes in the photoelectron current. We can use both the sweep data and the fixed bias potential measurements as long as the probe potential is negative. The setting of the fixed bias potential changed throughout the mission and was at negative values more often toward the later part of the mission. We could in principle use both of the two LAP probes for this study, but LAP2 suffered from a probable surface contamination \citep{johansson2017} for large parts of the mission. Moreover, LAP1 was sunlit most of the time and LAP2 was more often in shadow, and we therefore only use LAP1 (hereafter LAP) in this paper.

To measure the plasma density, we used the Mutual Impedence Probe  \citep[MIP,][]{trotignon2006}, which more accurately measures the absolute value of the electron density than the LAP instrument. MIP is not always capable of measuring the density when the density is too low. In particular, MIP suffers from a limiting lower threshold in the plasma density below which it cannot measure. This depends on the operational mode used, however.

\section{Measurements of flares}
\begin{figure*}
            \includegraphics[width=\hsize]{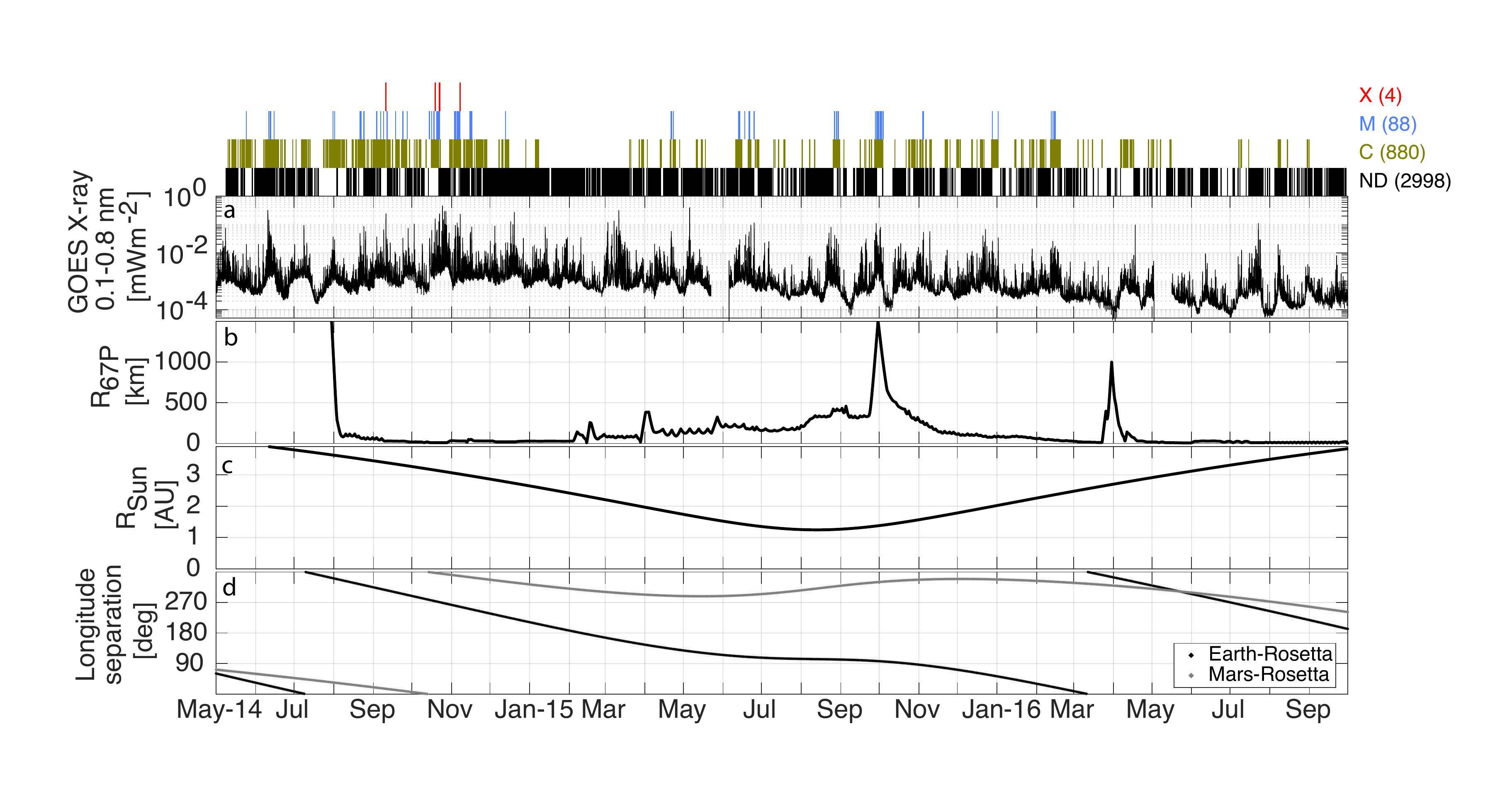}
        \caption{Time series of (a) GOES measured X-ray irradiance (were each peak corresponds to a flare), (b) Rosetta's distance to the comet nucleus, (c) the heliocentric distance of Rosetta, and (d) the heliospheric longitudinal separation between Rosetta and Earth as well as between Rosetta and Mars. The colored vertical lines at the top indicate the cataloged C, M, and X class flares that were observed at Earth and occurred during the Rosetta mission, which were also directed such that Rosetta could observe them. Those that were not detectable (ND) because of the geometry (too far out on the limb of the Sun) are indicated by the black lines.}
 \label{overview}
\end{figure*}

The Rosetta spacecraft followed comet 67P closely for more than two years, from 2014 to 2016. During this interval, the GOES spacecraft at Earth recorded about 4500 solar flares, observed in X-ray, of classes B to X. The X-ray fluxes measured by GOES during the Rosetta mission are plotted in Figure \ref{overview}a. 
These flares were automatically detected, categorized, and cataloged. We used the Hinode Flare Catalog  (\url{hinode.isee.nagoya-u.ac.jp/flare_catalogue}), which has easily retrievable information on the flare location on the Sun (latitude and longitude) \citep{Watanabe2012}. The majority of these flares were not observable since Rosetta and Earth were separated too far in heliocentric longitude for a large portion of the two-year interval. The longitudinal separation between the Earth and Rosetta is shown in panel d. For example, a separation of 180$^\circ$ means that Rosetta and Earth are on opposite sides of the Sun, and a separation of 0$^\circ$ or 360$^\circ$ means that they are at the same longitude. Rosetta and Mars were closer in longitude more often than Rosetta and Earth during this interval. From the information on the solar flares' location on the Sun we can determine which of them could be observed by Rosetta from a geometrical point of view, assuming that flares radiate isotropically, but not through the Sun itself. To minimize any limb-darkening effect, we set an arbitrarily chosen limit that Rosetta needs to be within a solar longitude of $\pm70^{\circ}$ from the flare location. We found that a total of 1504 flares of all classes could be observed. We set no restriction on the flare location on the Sun as seen from Earth to allow as many solar flares as possible to be included in our study. However, we can note that when the same $\pm70^{\circ}$ limit is set on the flare location on the Sun as seen from Earth, the number of observable flares is reduced to 997, but a few flares are then also missed that have a clear effect in Rosetta data (see the information on location in Col. 3 of Table \ref{fla}).

 Throughout large parts of the Rosetta mission, of course flares occurred on longitudes on the Sun that can not be observed from Earth, but could be observed by Rosetta. Below we introduce measurements from Mars, which can remedy this situation somehwat. The flares of classes C, M, or X that are potentially observable by Rosetta are indicated by vertical lines at the top of Figure \ref{overview}a, together with the non-detectable (ND) flares of the same classes. Geometrically detectable B- class flares are not included in Figure \ref{overview} because they are too numerous, but they are included in our further analysis. Flares of class A are not considered in this paper because they are simply too weak to have an effect, as we realized in a by-eye inspection of several events. As we show later, not even the B-class flares are intense enough in EUV irradiance to cause an observable effect in Rosetta data.

While X-ray flares are important for planetary atmospheres, this does not hold for the cometary coma of 67P, which is optically thin at these wavelengths. Similarly, the material on the Langmuir probe (TiN) provides a significant photoelectron yield for EUV wavelengths, but not for X-rays \citep{johansson2017}. It is therefore more important to obtain an accurate measure of the EUV irradiance during a flare, rather than the X-ray components. Adequate measurements of the full EUV spectra at a high cadence are, however, not always available. Below we describe the various measurements and estimates of the EUV spectra that are of interest.

The Solar EUV Experiment on the Thermosphere Ionosphere Mesosphere Energetics and Dynamics mission \citep[TIMED/SEE, ][]{woods2005} measures the EUV irradiance, but high-cadence data are only obtained during 3 min every orbit around Earth ($\sim$100 min), which captures some flares, but not all. The Solar Dynamics Observatory carries the Extreme Ultraviolet Experiment (SDO/EVE), whose EUV measurement cadence and spectral resolution are high enough, but only during 3h per day for the 34-106 nm channel, while the 6-37 nm wavelength channel made measurements continuously. In addition, after 2014, SDO could not measure short-ward of 34 nm as a result of an instrument failure. The Solar and Heliospheric Observatory (SOHO) mission, in orbit around Earth's L1 Lagrange point, carries the Solar EUV Monitor (SEM) instrument, which measures the solar flux at high time-resolution (1 min), but only at certain wavelengths \citep{judge1998}. It can hence be used to confirm the presence of a flare and the timing of the EUV peak at that wavelength, but cannot determine the total irradiance. 

To compensate for the lack of adequate high time-resolution, continuous monitoring, and high spectral resolution EUV measurements, we used the flare irradiance spectral model version 1 \citep[FISM v.1,][]{chamberlin2008}. FISM v.1 is an empirical model that uses the GOES 0.1-0.8 nm irradiance measurements and their time-derivative as proxy for flares and is also calibrated to measurements from TIMED/SEE and the Solar Stellar Irradiance Comparison Experiment (SOLSTICE) on the Upper Atmospheric Research Satellite (UARS) for some flares that were captured during the high-cadence measurements \citep{chamberlin2008}. For a full description we refer to that paper, but we can only mention that the model provides continuous estimates of the irradiance in the range 0.1-190 nm at a 1 min resolution. The accuracy of the FISM v.1 model is wavelength dependent, but the error in estimated irradiance is at least lower than 40\% above 14 nm \citep{chamberlin2008}. 

The Mars Atmosphere and Volatile Evolution (MAVEN) mission has been in orbit around Mars since late 2014 and beyond the mission end of Rosetta. As Mars and Rosetta followed each other in solar longitudes during large parts of the Rosetta mission, MAVEN can provide some additional measurements of flares that are detectable at 67P. MAVEN carries the EUV Monitor (EUVM), which measures at a cadence that is high enough, but only at certain wavelength bands \citep[0.1-7 nm, 17-22 nm, and 121.6 nm, see][]{eparvier2015}. Because the MAVEN spacecraft is in an elliptical orbit around Mars and therefore regularly dives into the Martian ionosphere and upper atmosphere, as well as intermittently shifting EUVM's pointing away from the Sun, it only measures the solar EUV flux $\sim$63\% of the time. Similar to the FISM v.1 model, the FISM-Mars (FISM-M) model was developed. FISM-M is based on the EUVM measurements from MAVEN and is calibrated against flares captured by SDO/EVE. Similar to its predecessor FISM v.1, FISM-M provides estimates of the EUV irradiance during solar flares at a 1 min cadence and in the interval 1-190 nm \citep{thiemann2017b}.
Furthermore, the Stereo-A satellite is capable of EUV imaging and thereby detecting flares, but the flare catalog generated by that mission ended in 2012 and will not be updated in the near future (M. Aschwanden, private communication). 
In summary, by combining the FISM v.1 and FISM-M estimated EUV irradiance with X-ray measurements from GOES and EUV measurements from SOHO and MAVEN, we can reliably determine which flares Rosetta was exposed to, and what their intensities were. The data from TIMED/SEE, SDO/EVE, and FISM v.1 can be found at \url{http://lasp.colorado.edu/lisird/}, the SOHO/SEM data at \url{https://dornsifecms.usc.edu/space-sciences-center/} , and the FISM-M data are available through the NASA Planetary Data System (PDS) \url{https://pds-ppi.igpp.ucla.edu/search/?sc=MAVEN&i=EUV}.

As Rosetta followed the comet, it traveled from a heliocentric distance of about 4 AU, passed perihelion at 1.25 AU, and proceeded outward in the solar system to about 4 AU again (Figure \ref{overview}b). We thus have to correct the irradiances in the FISM models for the change in heliocentric distance, assuming a simple 1/r$^2$ relation (from Earth to Rosetta for FISM v.1 and from Mars to Rosetta for FISM-M). At the same time, the trajectory of Rosetta spanned distances of $\sim$10-1500 km to the nucleus, as is shown in Figure \ref{overview}c. The observed neutral density and the level of dynamics in the plasma environment at the location of Rosetta thus vary with both heliocentric distance and cometocentric distance, and affect how easy or difficult it is to detect a flare in the Rosetta LAP data, which we discuss further below.

\begin{figure*}
            \includegraphics[width=\hsize]{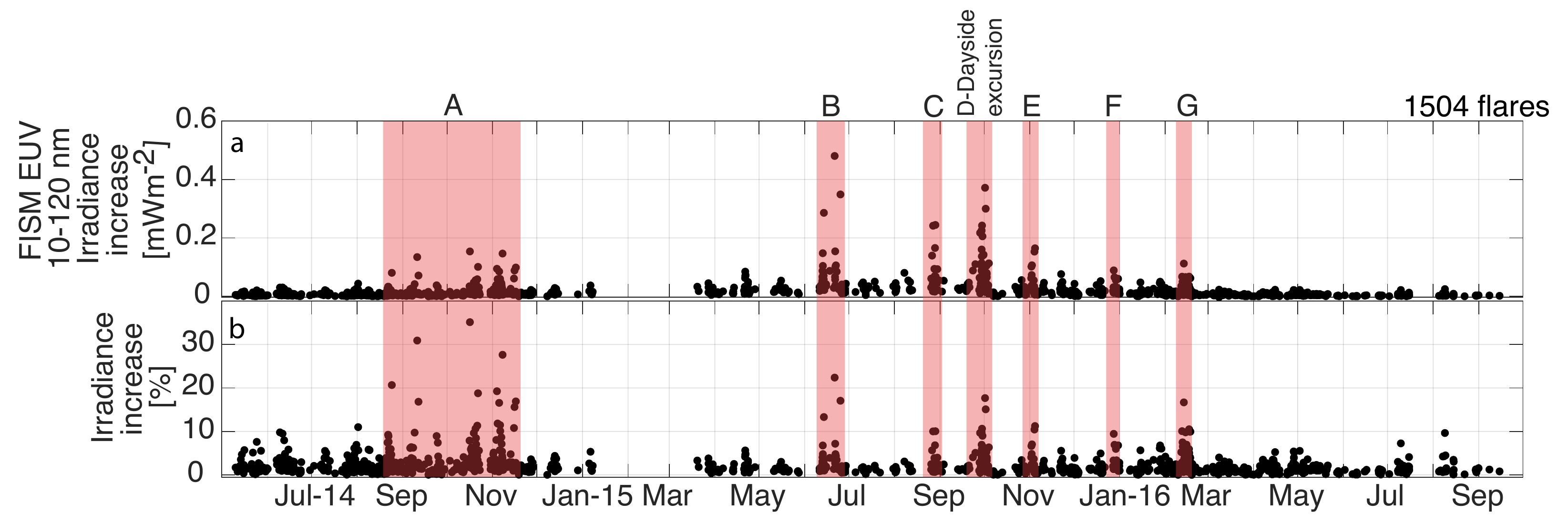}
        \caption{Increase in EUV irradiances (10-120 nm) for all Rosetta-directed flares plotted in (a) absolute values and (b) as percentage changes. The irradiance is estimated in the FISM v.1 model. Seven intervals (A-G) are highlighted with flares of high increases in EUV irradiance.}
 \label{inc}
\end{figure*}

\begin{figure*}
            \includegraphics[width=9cm]{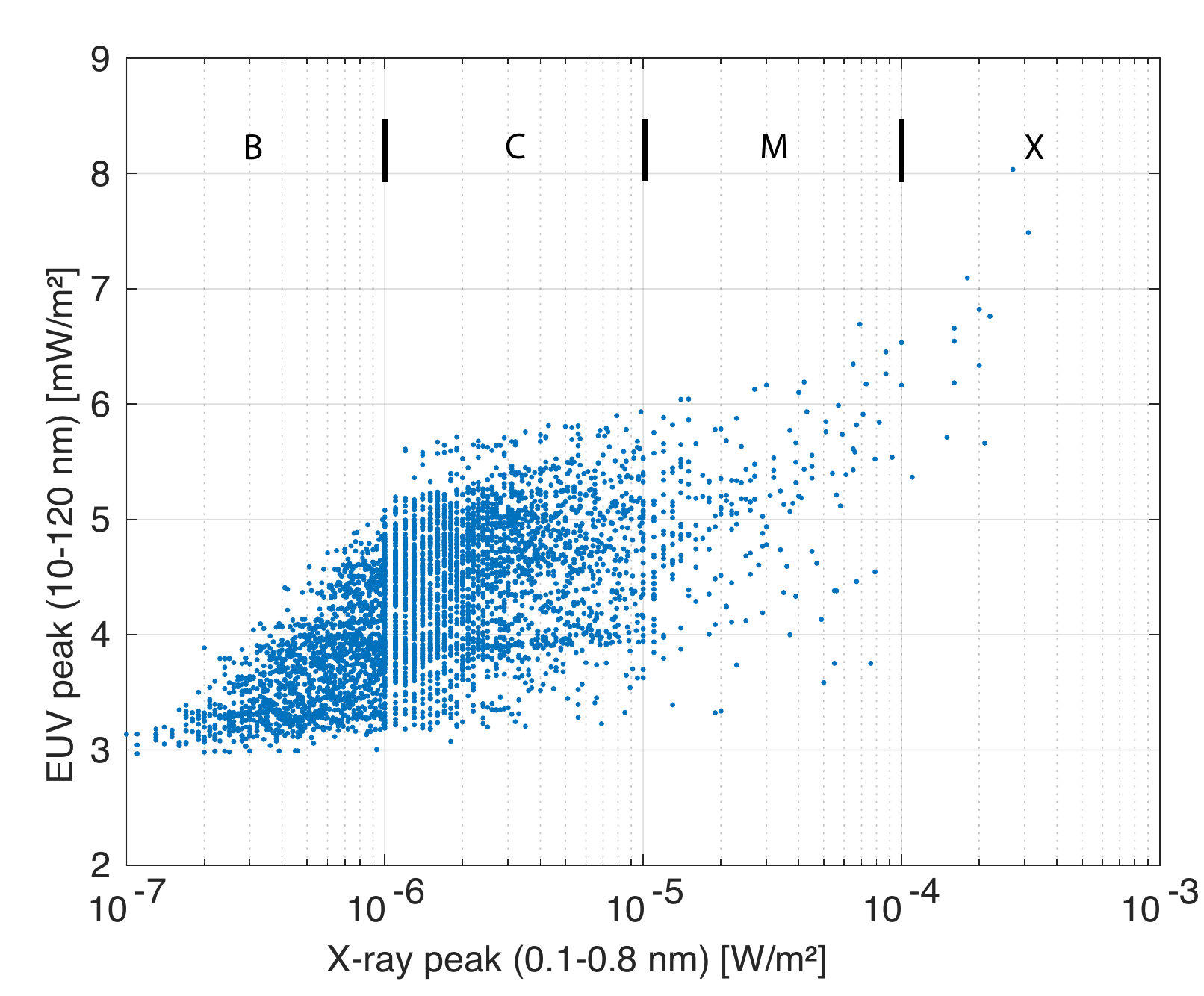}
        \caption{Peak of X-ray irradiances vs. peak of EUV irradiances for 4502 flares in the interval 2014-2016 when Rosetta was at comet 67P. The X-ray irradiances are measured by GOES, and the EUV irradiances are estimated by the FISM v.1-model. Not all of these flares are directed toward Rosetta.}
 \label{euvxray}
\end{figure*}

In Figure \ref{inc}a we show the increase in the FISM v.1 solar irradiance, corrected for heliocentric distance, for each of the 1504 Earth- and Rosetta-directed flares. The increase is calculated as the peak value of the irradiance during the flare minus the 5min average immediately before the flare was initiated. Panel b shows the percentage change of the irradiance during each flare. We have highlighted seven intervals (A-G) when several flares with large increases in EUV irradiance occurred. 
When Figure \ref{overview} and \ref{inc} are compared, it can be seen that the largest increases in the X-ray irradiance do not necessarily correspond to the largest increases in EUV irradiance, although there is a relation. It is therefore not very meaningful to use the GOES A, B, C, M, and X categories for studies of flares when the EUV wavelengths are of importance, such as for comets and photoelectron currents of Langmuir probes. Figure \ref{euvxray} shows the relation between the X-ray (from GOES) and EUV irradiances (from FISM v.1) for all flares that occurred during the Rosetta mission. A trend is visible (EUV $\propto$ log(X-ray)), but the scatter is also large enough to make the predictions of how the EUV changes depending on the X-rays uncertain.

\section{Flares observed by Rosetta}
A first simple by-eye inspection was conducted of all 1504 flares directed toward Rosetta to determine which effects the flares might have on both the plasma density and the LAP photoelectron current. This revealed that few flares showed any clear, or large, effects at all. No obvious plasma density increases were found, and only minor changes to the photoelectron current were observed in conjunction with most flares. In the following section we therefore show several detailed examples of when we did see effects and examples of when we did not see any clear effects. All flares that we did see effects from in Rosetta data occurred during the highlighted intervals in Figure \ref{inc}, which accentuates that the increase in EUV flux is of importance (the intervals were selected by eye based on the high-percentage increase of some of the flares within those intervals). On the other hand, not all high EUV flares are observed to cause any effects at 67P and Rosetta. The presented cases represent the clearest and largest effects that we were able to find in our survey.

\begin{figure*}
            \includegraphics[width=\hsize]{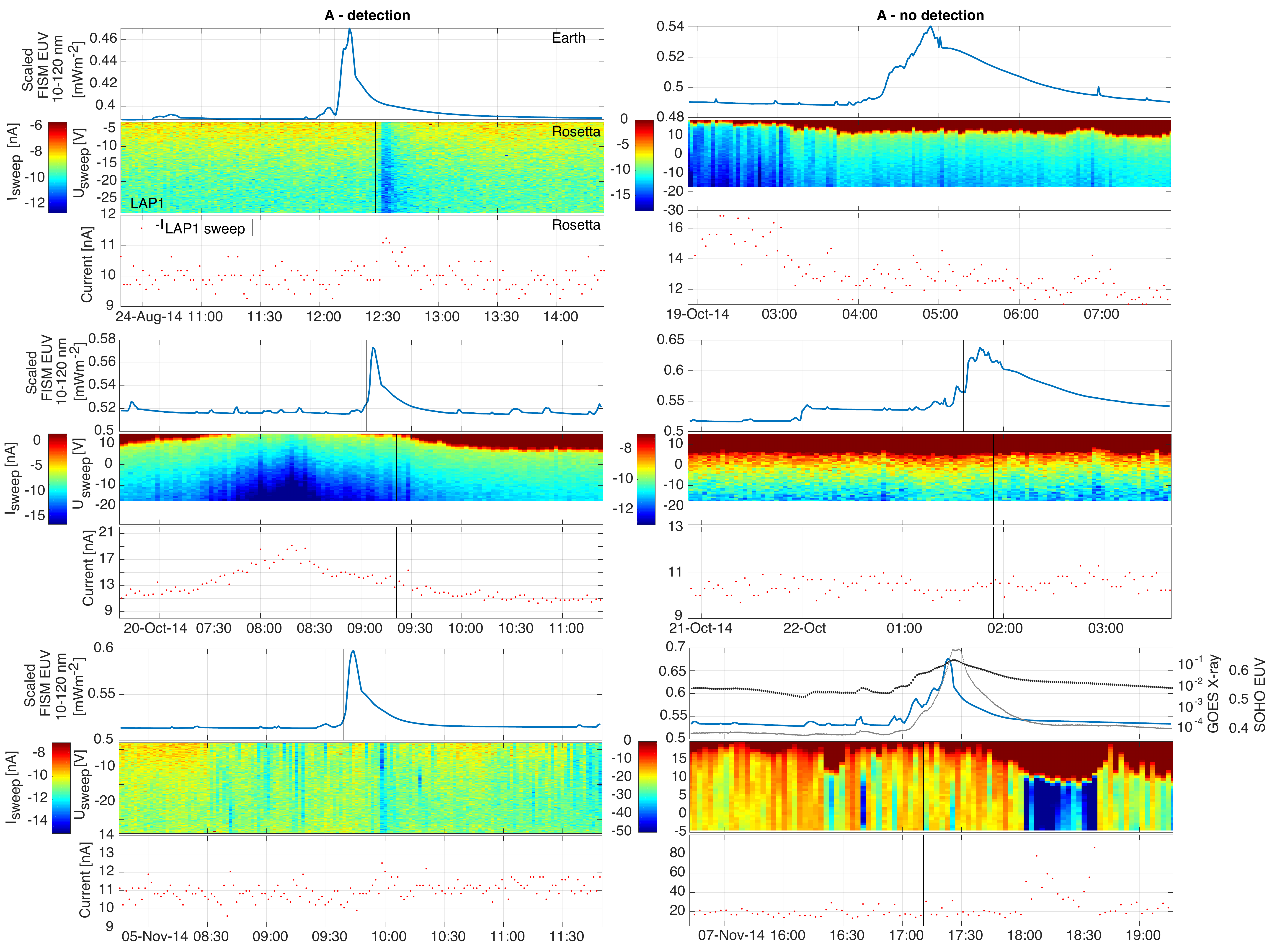}
        \caption{Six examples of flares and their effects, and lack of observable effects, in Rosetta data. In each of the six cases the top panels show FISM v.1 irradiance, and the middle panels show LAP sweeps, where the collected current is color-coded and the applied bias voltage is indicated on the vertical axis. The lower panel shows the current from the maximum negative bias voltage from the sweeps. A change in current is due to a change in either plasma density or photoelectron emission. The black vertical line in the top panel indicates the onset of the flares at Earth, and in the lower two panels, this line has been time-shifted by the speed of light travel time from Earth to Rosetta. The scales on the axes change between all six examples. In the lower right example we have added the X-ray irradiance at 0.1-0.8 nm from GOES (black, in units of $10^{10}$cm$^{-2}$s$^{-1}$) as well as EUV fluxes at 1-50 nm from SOHO (gray, in units of mWm$^{-2}$) to show an example of the timing of the same event in different wavelength bands.}
 \label{areaa}
\end{figure*}

Figure \ref{areaa} shows six examples of flares and their effect in Rosetta data from interval A. By eye, we identified effects in the photoelectron current in the three cases to the left and see no clear effects in the three cases to the right. The top panels show the FISM v.1 data with the magnitude scaled but not time-shifted from Earth to Rosetta. The onset of each flare at Earth is indicated by the black vertical line. The second panels show LAP sweep measurements from mainly the negative biased part, that is, when the probe potential is biased to negative voltages such that positive ions are collected and electrons, including emitted photoelectrons, repelled. We recall that a change to higher negative values (darker blue) in the collected current means an increase of emitted electrons (or attracted ions). In the lower panels we plot the negative of this current, at the maximum negative bias voltage during the specific interval. The maximum negative bias voltage is often -30 V, but in the top two cases to the right in Figure \ref{areaa}, it only reaches -18V, for instance. The fixed bias voltage was set to positive values in these intervals and therefore was not useful for tracking the photoemission. The scales on the axes are shifted between each case. Keeping the same scale is not possible because the overall plasma conditions change throughout the mission.
The black vertical lines in panels 2 and 3 indicate the expected arrival of each flare when it is time-shifted by the Earth-Rosetta distance divided by the speed of light. This time shift varies between about 2 and 40 min throughout the mission, depending on the relative positions of Earth and Rosetta. 

The examples in the left column in Figure \ref{areaa} show that at the expected arrival time of the solar flares, an increase in the LAP photoelectron current is indeed observed. The collected current increases by typically a few nA for an EUV increase of $\sim$0.1 mW/m$^2$. The effect is hence quite modest. The effect is clearest in the uppermost example and less so in the lower two examples. The observed increase in current is shorter in time than the duration of the flare, which can be interpreted as that the EUV irradiance apparently needs to overcome a certain threshold value to cause an effect. However, this is not very clear, and considering the overall uncertainties, this cannot be  determined accurately.  To the right, three flares with apparently similar EUV intensities show no apparent increase in the photoelectron current that stands out from the overall fluctuations. The top right example, from 19 October 2014, is an X-class flare that was seen to make a significant effect at Mars \citep{thiemann2015,peterson2016}. At the comet, however, it is hard to distinguish any increase in the measured current (or density from LAP - not shown) that stands out from the rest of the variations. There is a possible peak just after 05:00 UT, but it is very narrow compared to the broad flare and occurs before the main peak of the flare would have arrived at 67P. There are also similar sized peaks at 04:40 and 07:00 UT, which makes us reluctant to identify this as an effect of the flare. However, it could still be that the peak just after 05:00 UT is an effect of the flare, but it does in any case not stand out in relation to other variations in the plasma environment. There are also errors in the timing of the FISM v.1 model, which could explain why the current peak occurs before the expected flare peak. 

In the lower right two examples it is even harder to find any response to the flare within; the reason is somewhat unclear. The LAP probe is not in shadow, it is operated in a similar mode, the spacecraft potential is still negative, and the flares occurred on solar longitudes in view of Rosetta (although some limb darkening could still occur). The lower right example is interesting because of the large effect seen at 18:00-18:30 UT, which is of similar duration as the flare seen in the FISM v.1 data. This is, however, too late to be caused by the solar flare itself if the timing is correct. The explanation for this particular signature in LAP data rather seems to be the solar wind CIR that impacted on the comet \citep{edberg2016}, which caused generally large plasma disturbances (higher density and more fluctuations). The eventual solar flare signature could simply have drowned in the otherwise dynamic plasma interaction. We show additional examples like this in the next figures. 

The timing of the flares in the FISM v.1 flare model is based on GOES X-ray data, and the EUV irradiance peak of the flare can occur several tens of minutes after the X-ray peak \citep{thiemann2017}. We therefore checked the timing of our events using SOHO EUV and SDO/EVE EUV high-cadence measurements. For the event from 7 November 2014 (lower right in Figure \ref{areaa}), the time difference between the FISM v.1 EUV peak, the SOHO peak, and the X-ray peak are no more than a few minutes, as the plot shows. We also checked the timing of all other examples, which we  presented below, and we browsed through all of the 1504 events to make sure that we searched for signatures at the correct time.

\begin{figure*}
            \includegraphics[width=\hsize]{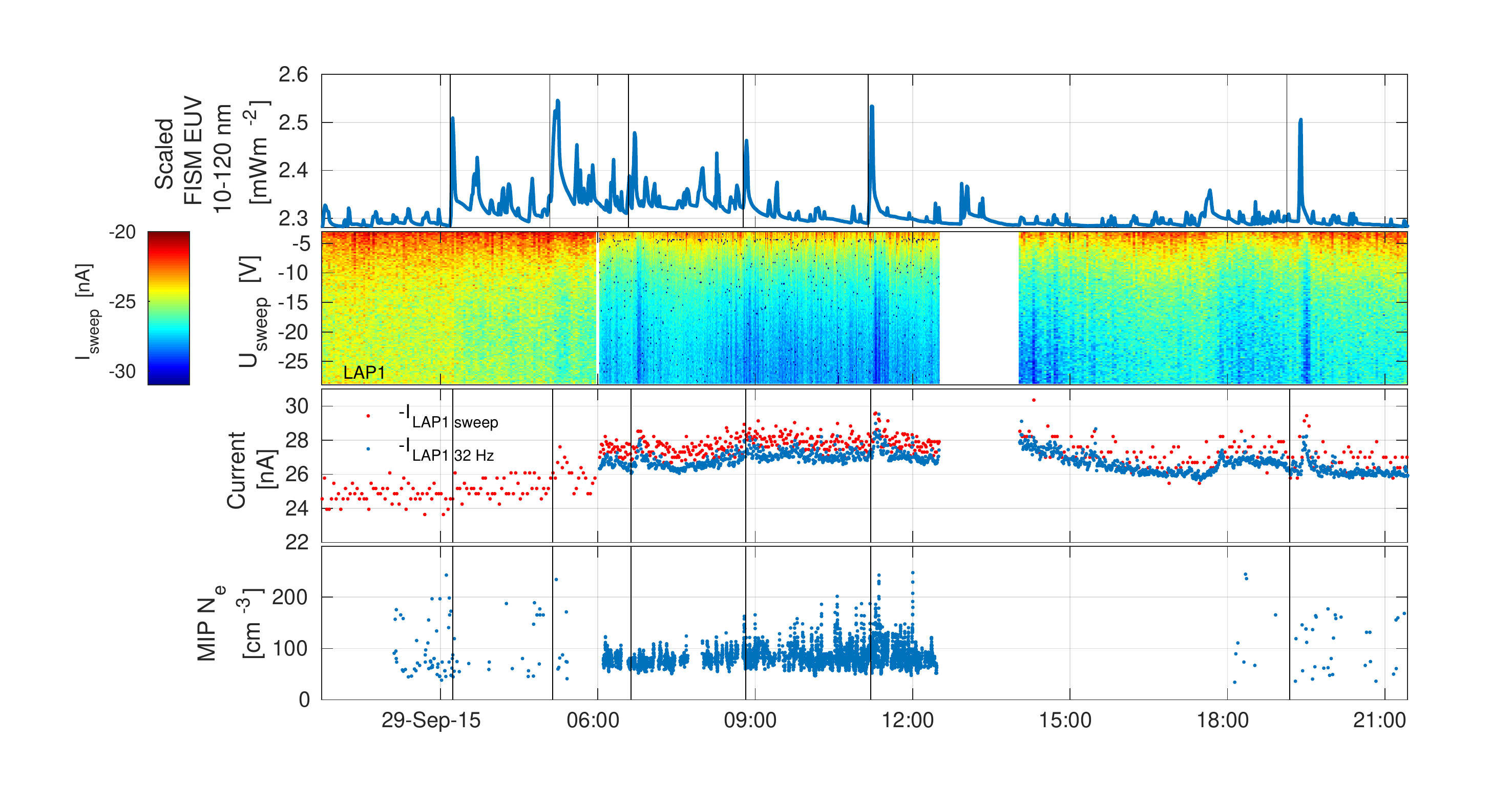}
        \caption{Examples of flares during the Rosetta dayside excursion (interval D in Figure \ref{excur}). The format is the same as in the previous Figure \ref{areaa}, but an additional panel is added showing the electron density as measured by the MIP instrument on Rosetta. In the third panel, we also show the current measured at a fixed negative bias voltage of -30V, which was available intermittently in this interval. Six flares are highlighted by black vertical lines, and all of them cause an increase in the measured current. Any increase in plasma density due to the flares is challenging to separate from the overall variability.}
 \label{excur}
\end{figure*}

From inspection of interval A, we direct the focus to interval D since that occurred during an interval when Rosetta was at a distance of up to 1500 km from the nucleus (known as the "dayside excursion" \citep{mandt2016}) and the local neutral and plasma density was relatively low. Fortuitously, a burst of intense flares occurred in this interval, which is shown in Figure \ref{excur}.
From this interval, the six largest flares are indicated by black vertical lines. Upon arrival of the flares at the comet, the current measured by LAP clearly increased for all of these events. The current increase is seen in the sweep measurements in panel 2 as the color becomes darker blue, as well as in panel 3. For this interval we also measured the ion current when it was at a fixed negative bias voltage (blue dots). These blue dots have a higher time-resolution (downsampled to 32 s shown here) and capture the flares better. A clear peak is seen at the arrival of each flare. The first two flares in this interval, at about 03:00 and 05:00 UT, are harder to identify since the LAP instrument was in a different mode with lower time resolution, and it did not measure at a fixed negative bias voltage. The sweeps indicate a possible jump of about 1-2 nA, slightly above the level of the overall fluctuations, when these flares hit. The next four flares indicted by the black vertical lines are somewhat clearer for identifying effects, and the 1-2 nA increases stand out, especially in the ion current (blue dots). However, there are also peaks in the LAP current that are not connected to a flare reported by FISM v.1 (e.g., at about 14:40 and 15:20 UT), and some smaller flares exist that are not indicated by black vertical lines but may also cause an increase in the photoelectron current (e.g., at 06:20 and 09:30 UT). The flare at 17:45 UT occured during a spacecraft attitude manoeuvre, which could affect the plasma density measurement, and was therefore disregarded.

\begin{figure*}
            \includegraphics[width=\hsize]{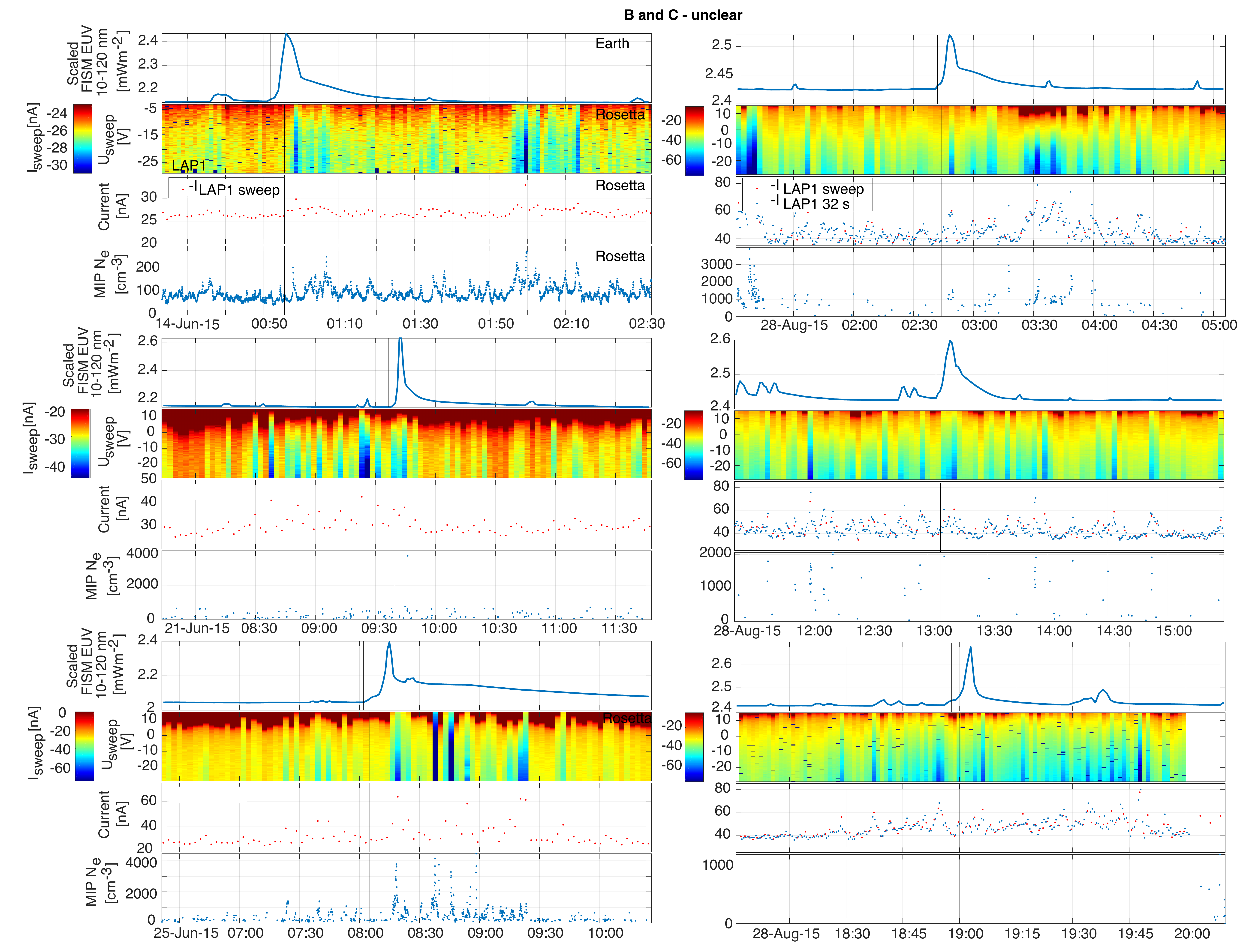}
        \caption{Six additional examples of flares from interval B and C in Figure \ref{overview}. The format is the same as in Figure \ref{areaa}, but we have added a fourth panel to each example showing the electron density measured by MIP, which was available or these intervals.}
 \label{areab}
\end{figure*}

Figure \ref{areab} shows in the same way as in the previous Figure \ref{areaa} six further examples of flares and their possible signatures in Rosetta data. These events are now from intervals B and C in Figure \ref{inc}. At this time, 67P was closer to perihelion and therefore more active. This is visible by the more variable and higher currents measured by LAP throughout the interval, for example. Fluctuations on the order of 10 nA are common in the data, which are much higher than the effects that flares caused during the less active phase. In these intervals we can only with difficulty pick out these six examples as the best cases for possible flare effects. Following each flare, there is a sudden increase observed in the measured current, but the difficulty is of course to separate them from the signatures of an otherwise dynamic plasma. From the positive detection of flares in Figure \ref{areaa} we can assume that the current increase due to similarly sized flares should be approximately the same, but for most cases in these six examples, the current increase is larger than in the previous examples, $\sim$10 nA compared to $\sim$1 nA. This could mean that the flare also causes an increase in the plasma density. The electron density is measured by MIP and included in the fourth panel. We recall that the current measured by LAP is the sum of all possible contributing currents, such as the photoelectron current and the ion current from the ambient plasma. There is typically an increase in the MIP density during all events in Figure \ref{areab}. However, there are also similarly sized density increases before as well as after, which are not linked to any flare. This makes it difficult to argue that the density increases are caused by flares rather than unrelated plasma dynamics. The large number of flares makes it somewhat probable that some flares would occur coincident with a density increase but that there is no clear causal connection. There are also flares in this interval of the same size, but they cannot be linked to any density increase. The lower left event stands out from the rest as there are large density fluctuations after the flare, which apparently start to appear upon impact. This flare also lasts longer than average. 

\begin{figure*}
            \includegraphics[width=\hsize]{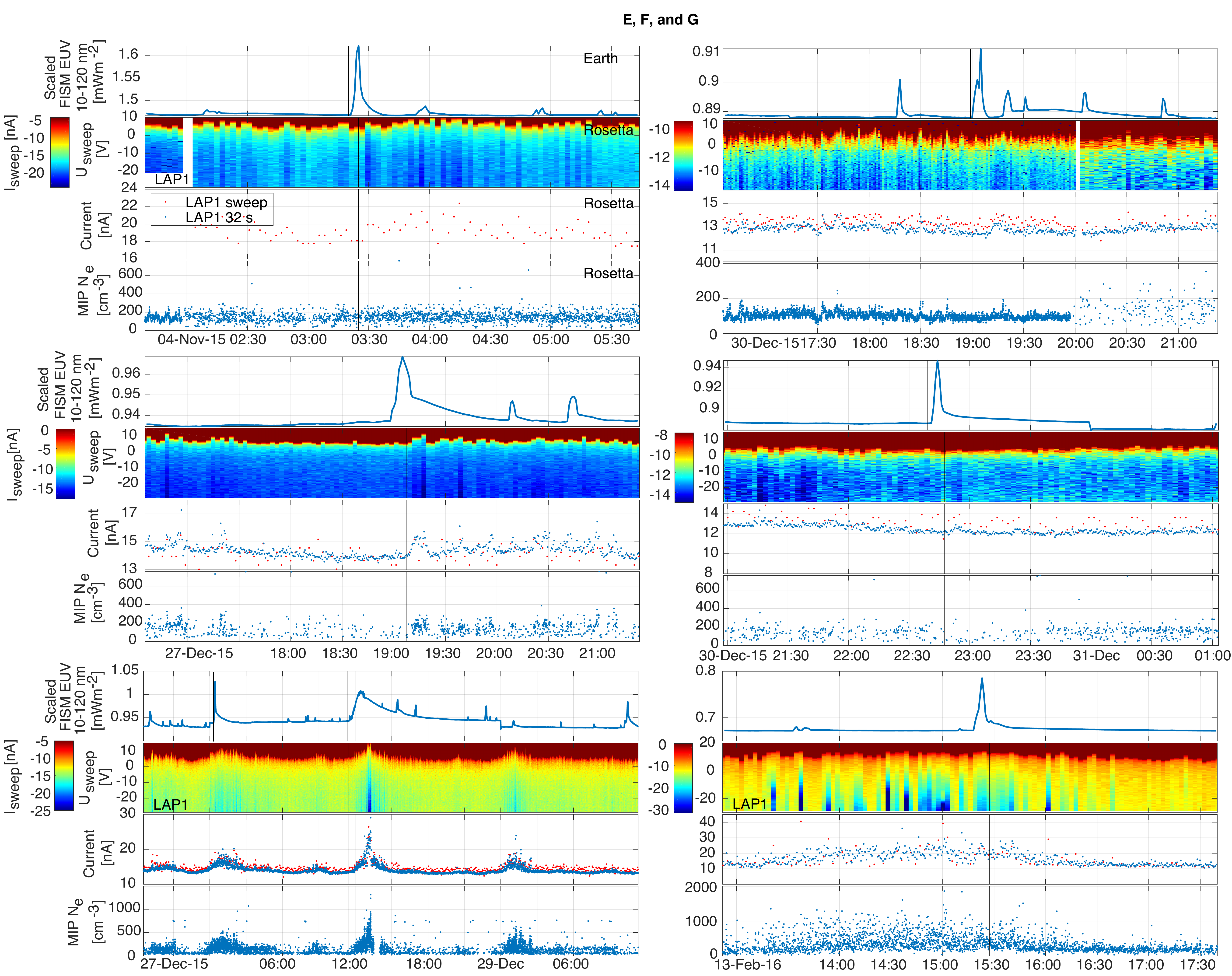}
        \caption{Seven additional examples of flares from interval E, F, and G in Figure \ref{overview}. The format is the same as in the previous figures.}
 \label{areaefg}
\end{figure*}

Finally, Figure \ref{areaefg} shows the best events that we could select from intervals E, F, and G. Now the local plasma dynamics is again more modest and the signatures in the LAP current due to a flare are easier to pick out. The lower right example illustrates when it is questionable if there is an effect at all. There is a weak increase in the current at the right time at 15:40, but it does not stand out significantly from the overall variability. The lower left example is potentially more interesting. There is a large increase in both density and current following the two flares in this example. The plasma environment in this interval is strongly modulated by the rotation of the comet nucleus. As the comet rotates with a period of $\sim$12 h, it will face a more active region toward Rosetta twice per rotation, and the local plasma density consequently changes together with the neutral density (not shown) \citep{edberg2015, odelstad2015}. The two flares both occur during maximum of these plasma variations, separated by 12h. These two flares are similar in magnitude, but the first is considerably shorter than the second. The apparent response of the plasma environment is also quite different. The second flare seems to generate a current increase of $\sim$6 nA and the density increases by a factor 2, from 500 to 1000 cm$^{-3}$, while the first flare does not seem to cause any effect at all above the gradual change that is due to the nucleus rotation. The effect of the second flare is comparable to those presented in Figure \ref{areab} during the more active phase of the comet.

In Table \ref{fla} we have listed all 24 of the individual flares of which we observe effects. We also include the three flares that were shown in Figure \ref{areaa} and that did not give any noticeable effect. Not all events listed in the table are shown in plots in this paper. The increase in EUV irradiance and photoelectron current (I$_{ph0}$) is determined from the plots as the peak value minus the value immediately before the flare. For the photoelectron current, this is somewhat difficult in many cases because of the overall variability. The photoelectron current is estimated from either the sweep values or, when available, the current measured at a fixed bias-voltage of -30 V. The flares that were harder to separate from the ambient plasma variability, shown in Figure \ref{areab}, are indicated by italics. In this group we also include the bottom two examples from Figure \ref{areaefg}. 

\begin{table*}
\caption{24 flares observed to cause a noticeable effect at Rosetta, as well as 3 flares that did not show any effect. The information on the GOES class and the location on the Sun are taken from the Hinode flare catalog. The flares that were harder to separate from the ambient plasma variability are written in italics. The dashes indicate that no information is available (e.g., not seen at Mars in the FISM-M model, or that no effect was seen). The time information refers to the observations at the GOES satellite.}             
\label{fla}      
\centering          
\begin{tabular}{c c c c c c}     % 7 columns 
\hline\hline       
                      % To combine 4 columns into a single one 
Date and Time (UT) & GOES Class & Location & $\Delta$EUV [mW/m$^2$] & $\Delta$EUV [mW/m$^2$] & $\Delta$I$_\mathrm{ph0}$ [nA]\\ 
                     &      & on Sun & FISM v.1    & FISM-M      & LAP \\
\hline                    
   24 Aug 2014 12:00 & M5.7 & S09E76 & 0.08 & -    & 1.2 \\  
   17 Oct 2014 15:35 & C6.7 & S15E79 & 0.03 & -    & 1.5 \\
   19 Oct 2014 15:20 & X1.1 & S14E64 & 0.05 & 0.03 & -   \\
   20 Oct 2014 09:01 & M3.9 & S14E42 & 0.06 & 0.02 & 1.8 \\
   22 Oct 2014 01:17 & M8.7 & S13E21 & 0.09 & 0.03 & -   \\
   05 Nov 2014 09:26 & M7.9 & N20E68 & 0.08 & -    & 1.5 \\
   07 Nov 2014 16:53 & X1.6 & N14E36 & 0.15 & -    & -   \\
   14 Jun 2015 00:52 & M2.0 & N14W73 & 0.29 & 0.03 & 3.0 \\
 \it{21 Jun 2015 09:26} & \it{M3.8} & \it{S21W57} & \it{0.45} & -& \it{10}\\
 \it{25 Jun 2015 08:03} & \it{M7.9} & \it{N09W42} & \it{0.35} & -& \it{21}\\
 \it{28 Jun 2015 02:43} & \it{C8.0} & \it{S13W59} & \it{0.09} & -& \it{11}\\
 \it{28 Jun 2015 13:05} & \it{M2.2} & \it{S14W65} & \it{0.17} & -& \it{11}\\
 \it{28 Jun 2015 19:00} & \it{M2.1} & \it{S13W70} & \it{0.26} & -& \it{10}\\
   29 Sep 2015 03:11 & M1.2 & S08W78 & 0.22 & 0.05 & 1.1 \\
   29 Sep 2015 05:05 & M2.9 & S21W37 & 0.25 & 0.1  & 2.0 \\
   29 Sep 2015 06:35 & M1.4 & S20W34 & 0.16 & 0.02 & 2.0 \\
   29 Sep 2015 11:09 & M1.6 & S20W39 & 0.25 & 0.04 & 2.3 \\
   29 Sep 2015 19:30 & M1.1 & S20W36 & 0.21 & -& 2.2 \\
   30 Sep 2015 13:15 & M1.1 & S23W59 & 0.15 & 0.03 & 2.5 \\
   02 Oct 2015 00:06 & M5.5 & S19W67 & 0.30 & -& 3.7 \\
   04 Nov 2015 03:20 & M1.9 & N15W64 & 0.15 & -& 2.0 \\
   27 Dec 2015 18:59 & C2.8 & N01W01 & 0.04 & -& 1.6 \\
 \it{28 Dec 2015 11:22} & \it{M1.8} & \it{S23W11} & \it{0.07} & - & \it{6.0}\\  
   30 Dec 2015 19:05 & C1.1 & S22W50 & 0.03 & -& 1.2 \\
   30 Dec 2015 22:42 & C4.3 & S24W39 & 0.06 & 0.02 & 1.0 \\
 \it{13 Feb 2016 15:16} & \it{M1.8} & \it{N13W25} & \it{0.12} & \it{0.02} & \it{9.0}\\ 
 \it{14 Feb 2016 19:19} & \it{M1.0} & \it{N15W47} & \it{0.07} & \it{0.02} & \it{5.0}\\ 
\hline                  
\end{tabular}
\end{table*}

\begin{figure*}
            \includegraphics[width=\hsize]{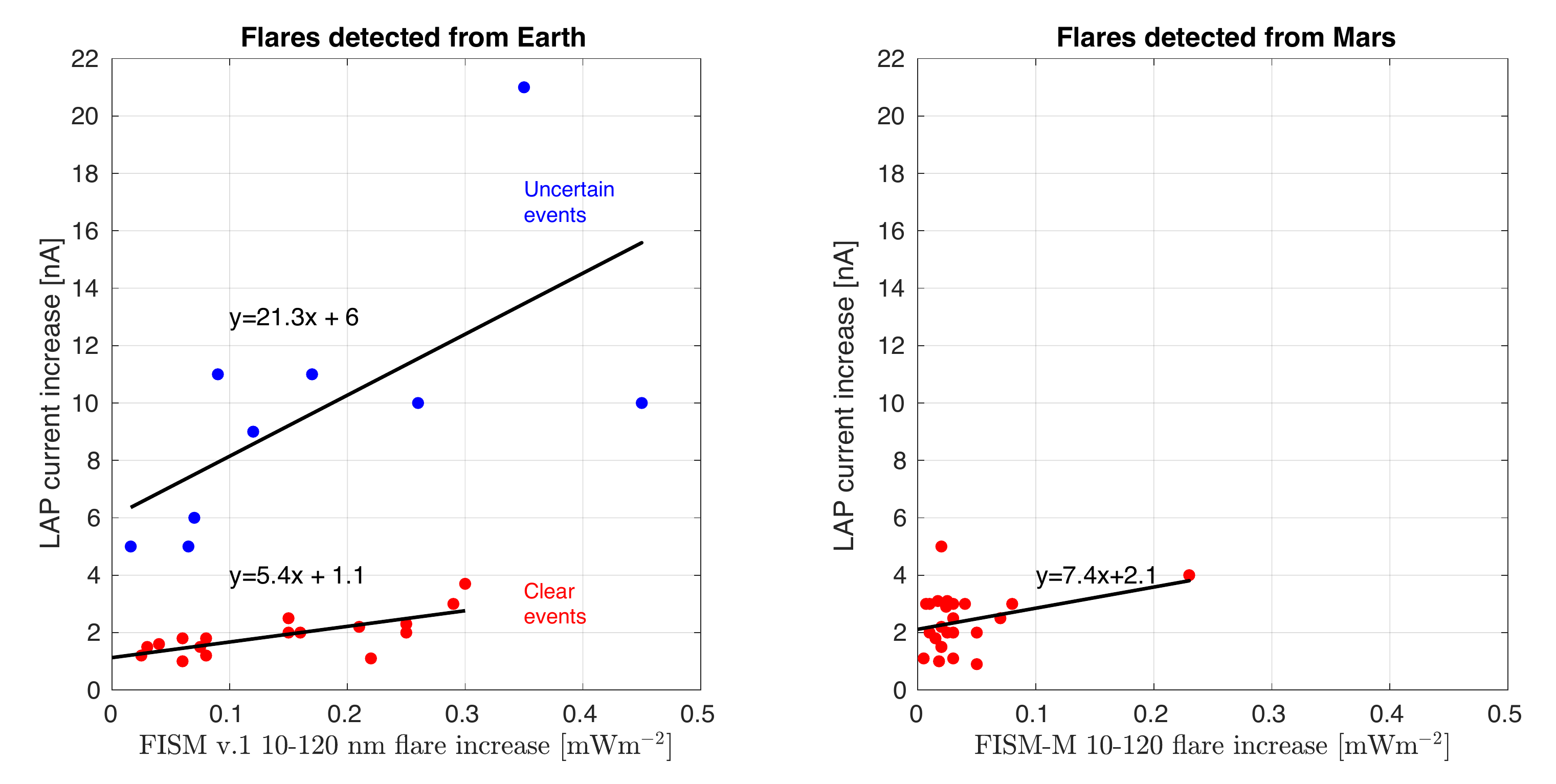}
        \caption{LAP current increase as a function of flare EUV irradiance increase for (left) solar flares identified from the vantage point of Earth and (right) for solar flares directed toward Mars. The red group in the left panel corresponds to the events that are easier to identify, while the blue group corresponds to some larger events that are harder to identify because the plasma environment is more dynamic. The upper line is a fit to all blue points. The error in the FISM v.1 and FISM-M EUV values is roughly 40$\%$ and the error in the LAP current increase is estimated to be of similar size. In the left panel, $R^2$ = 0.53 for the fit to the "clear" event and $R^2$ =  0.41 for the "unclear events". In the right panel $R^2 = 0.12$ for the fit.}
 \label{scatter}
\end{figure*}

The effects of flares on the photoelectron current from LAP are also shown in Figure \ref{scatter} (left panel) where the measured increase in LAP photoelectron current is plotted as a function of increased EUV irradiance from the FISM v.1 model. The right panel of Figure \ref{scatter} is discussed further below. The data points are divided into whether they were easily identified ("clear") or difficult ("uncertain"), corresponding to those written in upright font or italics in Table \ref{fla}. There is a trend for the clear events that the photoelectron current increases linearly with increasing EUV irradiance. A least-squares fit to the points is also included and is indicated by the black line. This fit does, however, not go through the origin, which one might expect. It is non-physical that a flare of 0 EUV irradiance increase would yield a photoelectron current increase at all, and the fit is clearly not perfect. The goodness of the fit, indicated by the coefficient of determination R$^2=0.53$, is rather poor and there is therefore little statistical evidence for such a linear relation. One plausible explanation for it not going through the origin is that our selection is biased for low EUV flares: when the photoelectron current increase is not large enough to take it above the level of the ambient plasma fluctuations, we fail to see it. This will cause an over-representation of flares with larger effects for low EUV values, and hence the fit has a smaller slope than it should. It could also be that the trend is not linear throughout the interval, or perhaps more likely, that there are errors in the estimate of both the photoelectron current and irradiance, which makes the fit uncertain: the error in the FISM v.1 EUV values is roughly 40$\%$ and the error in the LAP current increase is estimated to be of similar size. 

The uncertain events (blue dots) also show an approximately linear relationship between EUV and photoelectron increase, but with a higher offset in current and a larger spread in the data. The fitted line does not go through the origin in this case either and R$^2=0.41$ for this fit, making it quite uncertain. The error in estimating the increase in photoelectron current is harder in these cases because the overall variability in the plasma is larger. 

\section{Mars-directed flares}
\begin{figure*}
            \includegraphics[width=\hsize]{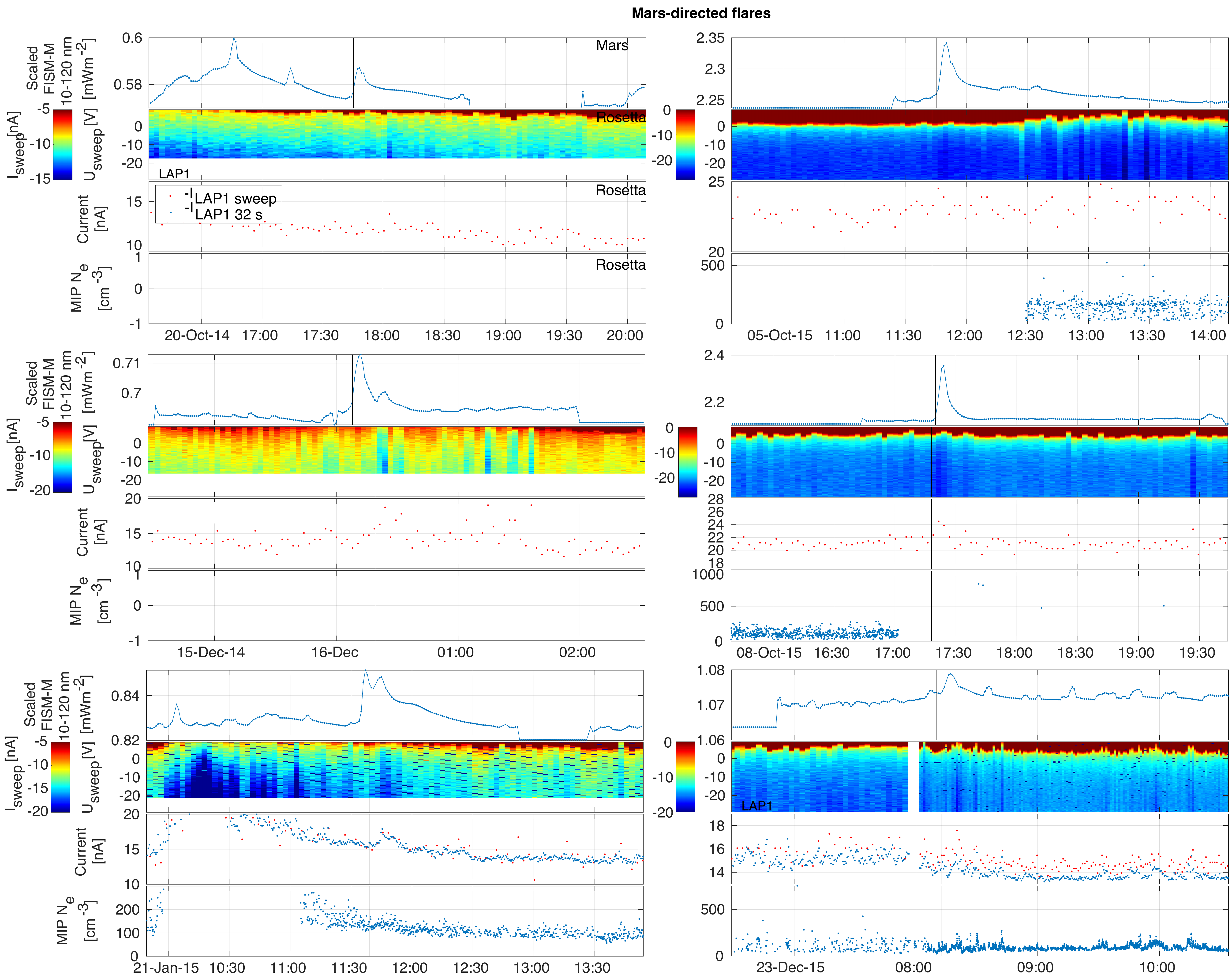}
        \caption{Same as Figures \ref{areaa}, \ref{areab}, and \ref{areaefg}, but for Mars-directed flares.}
 \label{areamars}
\end{figure*}

Because we have relatively few events that we could identify from Earth-directed flares, we also included Mars-directed flares in our study. At Mars, we only have measurements from the EUVM instrument on MAVEN, which is incorporated into the FISM-M model. We do not have information on the location on the Sun where the flares emanated. This means that when Rosetta and Mars are separated in heliospheric longitude, we cannot be certain that a flare that is seen at Mars also hits Rosetta. We therefore cannot obtain reliable statistics on what percentage of flares observed in FISM-M were also seen in Rosetta data. Furthermore, from FISM-M we only obtain irradiance data when the pointing of MAVEN is favorable and when it is not inside the induced magnetosphere of Mars. This causes some intermittency in the data and decreases the number of possible events that can be detected. A list of Mars-directed flares has been assembled by the MAVEN science team, but as this list is not comprehensive, we instead manually searched the FISM-M data for all possible flares. Using first an automatic peak finding algorithm, we found around 1000 peaks in the data. Not all of these were necessarily solar flares, but they might rather be any type of peak in the data (including stray data points). We then manually browsed through all events to identify flares that caused any visible effect in Rosetta data. Twenty-four events were found, coincidentally the same number as for the Earth-directed flare, which were similar to the events shown in Figures \ref{areaa}, \ref{areab}, and \ref{areaefg} in terms of increase in photoelectron current amplitude and duration and a negligible effect on the plasma density. These 24 Mars-directed flares are listed in Table \ref{flamars}, and 6 examples are shown in Figure \ref{areamars}. Some of them overlapped with the Earth-directed events found in FISM v.1 data and are also included in Table \ref{fla}. In the same way as before, we plot the increase in photoelectron current as a function of irradiance increase for all 24 events in Figure \ref{scatter} (right panel). A least-squares fit is added, which is similar to the Earth-directed flares, but the accuracy of the fit is too poor ($R^2 = 0.1$) for it to be meaningful. Very few flares with an irradiance increase above 0.1 mWm$^{-2}$ are found.  

\begin{table*}
\caption{Mars-directed flares that are observed to cause a noticeable effect at Rosetta. The stars indicate flares that are also observed by GOES and are too far out on the limb to be included in Table \ref{fla}.}             
\label{flamars}      
\centering          
\begin{tabular}{c c c c c c}     % 7 columns 
\hline\hline       
                      % To combine 4 columns into a single one 
Date and Time (UT) & GOES Class & Location & $\Delta$EUV [mW/m$^2$] & $\Delta$EUV [mW/m$^2$] & $\Delta$I$_\mathrm{ph0}$ [nA]\\ 
                     &      & on Sun & FISM v.1    & FISM-M   & LAP \\
\hline                    
 19 Oct 2014 02:15 & C3.2 & S17E46 & *    & 0.01 & 1.1 \\
 20 Oct 2014 04:35 & -    &  -     & -    & 0.02 & 1.5 \\
 20 Oct 2014 09:01 & M3.9 & S14E42 & 0.06 & 0.02 & 1.8 \\
 20 Oct 2014 17:45 & -    & -      & -    & 0.01 & 1.8 \\
 16 Dec 2014 00:08 & -    & -      & -    & 0.02 & 5.0 \\
 30 Dec 2014 03:30 & -    & -      & -    & 0.02 & 3.1 \\
 03 Jan 2015 17:30 & -    & -      & -    & 0.02 & 1.5 \\
 09 Jan 2015 05:45 & C3.4 & S17E64 & *    & 0.01 & 3.0 \\
 21 Jan 2015 04:30 & -    & -      & -    & 0.02 & 2.9 \\
 21 Jan 2015 11:30 & C9.9 & S11E89 & *    & 0.03 & 3.1 \\
 23 Jan 2015 18:15 & -    & -      & -    & 0.03 & 2.0 \\
 28 Feb 2015 21:15 & -    & -      & -    & 0.05 & 2.5 \\
 14 Jun 2015 00:52 & M2.0 & N14W73 & 0.29 & 0.03 & 3.0 \\
 27 Sep 2015 23:25 & C7.9 & S21W14 & *    & 0.03 & 1.1 \\
 29 Sep 2015 03:11 & M1.2 & S08W78 & 0.22 & 0.05 & 0.9 \\
 29 Sep 2015 05:05 & M2.9 & S21W37 & 0.25 & 0.10 & 2.0 \\
 29 Sep 2015 06:35 & M1.4 & S20W34 & 0.16 & 0.02 & 2.2 \\
 29 Sep 2015 08:00 & -    & -      & -    & 0.01 & 1.0 \\
 29 Sep 2015 11:09 & M1.6 & S20W39 & 0.25 & 0.04 & 2.3 \\
 30 Sep 2015 13:15 & M1.1 & S23W59 & 0.15 & 0.03 & 2.5 \\
 05 Oct 2015 11:45 & -    & -      & -    & 0.08 & 3.0 \\
 08 Oct 2015 17:20 & -    & -      & -    & 0.25 & 4.0 \\
 23 Dec 2015 08:00 & -    & -      & -    & 0.01 & 3.0 \\
 30 Dec 2015 22:42 & C4.3 & S24W39 & 0.06 & 0.02 & 1.0 \\

\hline                  
\end{tabular}
\end{table*}

%\begin{figure*}
%            \includegraphics[width=\hsize]{scatter_earth_mars.pdf}
%       \caption{Same as Figure \ref{scatter} but for Mars directed flares. {  $R^2 %= 0.12$ for this fit.}}
% \label{mars}
%\end{figure*}

\section{Discussion}
We find that solar flares generally have a weaker effect on the cometary coma of 67P than other variations in the plasma (caused by, e.g., variations in the neutral background, the amount of cold plasma present, plasma waves, and transient structures in the magnetic field). No clear effect of increased ionization and plasma density is found, although a few events do show some increase in plasma density in conjunction with solar flares. However, these are most often hard to distinguish from the overall plasma variability. One of the most promising events occurred on 28 December 2015 at around 12:00 UT when a solar flare impacted at the same time as a neutral and plasma density peak occurred (Figure \ref{areaefg}). Compared with similar density peaks 12 h earlier, which also coincided with a flare, or 12 h later, the plasma density was higher by a factor of 2 during this long-lived flare, even though the neutral density was similar to the one 12 h before and 12 h later, and the flux of high-energy electrons as measured by the Ion and Electron Sensor \citep[IES,][]{burch2007} did not show any increased values at this time (data not shown). The long duration should mean that the total amount of energy deposited is significantly higher, which could be the explanation for the apparent large effect.  However, from the total number of 1504 Earth-directed flares and several hundred Mars-directed flares we would by pure chance expect to see some flares during a coincidental plasma density increase, meaning that we cannot conclusively determine that the plasma density increase was caused by the flare in this case either. Larger flares than this are seen to not cause any noticeable effect. Rather than focusing on the peak EUV irradiance for each flare, one could also study the total amount generated by each flare by integrating over the duration of them. This would, however, not lead to any new detection of additional flare effects because we have already looked at all recorded flares throughout the mission. 

It should be noted that the likelihood of finding any effect is strongly dependent on the coma conditions. If there are large fluctuations in the plasma density caused by the general dynamic nature in the coma, it would not be possible to observe a single short-lived peak in density. If the coma had been significantly denser, such that also X-rays would be absorbed in the coma, we might have seen more effects of flares. We also note that some limb darkening of flares can occur, especially because we allowed flares at all longitudes as seen from the Sun to be included in our study. This effect would reduce the estimated EUV irradiance at Earth because the flare passes through the solar atmosphere, and  might lead to an underestimation of the flare intensity. This could add to some of the scatter in Figure \ref{scatter}, but because we find a similar trend from Mars-directed as from Earth-directed flares, this does not seem to be a main source of error.

The lack of clear evidence of plasma response to the increased EUV flux merits some further discussion. First, we may note that simulations and data alike suggest that the cometary (as opposed to solar wind) plasma dominates at the Rosetta position during most of the mission \citep[ e.g.]{edberg2015,yang2016,vigren2015, nilsson2017,heritier2017}. The observed plasma density followed the neutral gas density well \citep{odelstad2015,vigren2016b,galand2016}, but with much larger variations on timescales shorter than the nucleus spin period, presumably due to internal plasma dynamics as observed in simulations \citep{koenders2015,deca2017,huang2018}. This indicates that local (or close to local) ionization is important, and because the ionization source density (in units of m$^{-3}$s$^{-1}$) that is due to EUV should be proportional to the neutral gas density and to the EUV flux at relevant wavelengths, one might expect the dependence of the plasma density on EUV flux to be as strong as the previously mentioned studies found on neutral density. However, \citet{galand2016} found that high-energy plasma electrons could be more important ionization agents than EUV photons at low comet activity. This was followed up by \citet{heritier2018}, who showed in an extensive mission-wide study that EUV ionization dominated only in the months around perihelion (13 August 2015, at 1.25~AU). In our study, this corresponds to intervals B and C in Figure~2, where we have many fluctuations in the plasma and thus are often unable to determine if a flare contributes to the plasma variations. The absence of evidence for a correlation during the rest of the mission thus is consistent with the EUV flux being of minor importance for the plasma, in line with the above studies. However, it is interesting to note that this holds true for variations on the relatively short timescale considered here because it suggests that the flux of ionizing electrons is only weakly tied to the EUV flux, not acting only as a strongly amplifying secondary effect to primary ionization by solar EUV. 

While the effect on the plasma density is hard to observe, the solar flares have a clearer effect on the photoelectron current measured by the LAP instrument, at least for the few percent of all flares when the EUV irradiance increase is large enough. In 24 Earth-directed events (1.6\% out of 1504) and 24 Mars-directed flares with a higher-than-average increase in EUV irradiance we can see that the photoelectron current increases by a few nA, as shown in Figure \ref{scatter}. The photoelectron current increase is not far above the general "noise" level in the data, such that we generally require the EUV increase  to be large and the plasma conditions to be calm for the flare to be observable. The EUV increase during the flares is seldom above 10 \% (noting again that we used a model to obtain the EUV irradiance, which is only calibrated against some 30 flares and therefore not perfect). The EUV variation at Rosetta and 67P caused by the elliptic orbit of the comet around the Sun is on the order of 500 \% during the Rosetta mission \citep{johansson2017}, or in other words, it is  considerably larger.

On another topic, we may note that with a sudden change in the photoelectron current from the Rosetta spacecraft and the LAP instrument, we can assume that there will also be an increase in photoelectrons emitted from any dust particle in the coma. This will lead to an increased charge of that particle and thereby increase electrostatic forces within it. We can thus speculate that some of the flares might cause a fragmentation of dust particles as the electrostatic forces break up the dust grains into smaller particles. This might explain some of the increased variability in the plasma density seen after some flares (e.g., 25 June 2016 in Figure \ref{areab}). Such a speculative dust fragmentation could in turn lead to a decreased EUV flux in the coma as the increased number of particles would increase the total EUV absorption. The effectiveness of such a process is somewhat difficult to estimate and likely depends on a number of factors, such as the dust size distribution, their individual photoelectron yield function, charge state, solidity, density, and flow speed. We refrain from exploring this further in this paper and rather only mention it as an idea for further investigations in the future.

Although solar flares constitute a minor effect on the coma in comparison to solar wind CIRs and CMEs, for instance \citep{edberg2016, edberg2016b, hajra2018, noonan2018, goetz2018}, they will at least affect the entire coma at the same time because the coma is transparent to these wavelengths. This means that any effect seen locally at Rosetta can also be expected simultaneously in the entire coma, at least where the neutral, plasma, and dust properties are similar.

\section{Conclusions}
We find that solar flares have a weaker effect on the cometary plasma environment than other processes causing plasma variations, such as the solar wind and internal plasma processes. They do increase the photoelectron current from the LAP instrument, typically by 1-5 nA for flares with irradiance increases of up to 0.3 mWm$^{-2}$ in the wavelength interval 10-120 nm.
Solar flares are only detectable by LAP in 1.6\% of all cases (24 of the 1504 Earth directed flares) when plasma conditions are otherwise steady. Twenty-four Mars-directed flares were also detected by Rosetta. In a few cases the plasma density increases coincidentally with a flare in such a way that it is possible that the flare causes the density increase. It is generally difficult to distinguish effects from a flare from other variations in the plasma, however.

\begin{acknowledgements}
Rosetta is a European Space Agency (ESA) mission with contributions from its member states and the National Aeronautics and Space Administration (NASA). The work on RPC-LAP data was funded by the Swedish National Space Board under contracts 109/12 and 135/13 and Vetenskapsr\aa det under contracts 621-2013-4191 and 621-2014-5526. Work at LPC2E/CNRS was supported by CNES and by ANR under the financial agreement ANR-15-CE31-0009-01. We thank Matthieu Kretzschmar for useful discussions on solar flares.  CSW is supported by the Research Council of Norway grant no. 240000. EMBT's contribution was supported by the NASA MAVEN Program. The work of RH is financially supported by the Science \& Engineering Research Board (SERB), a statutory body of the Department of Science \& Technology (DST), Government of India through Ramanujan Fellowship at NARL. This work has made use of the AMDA and RPC Quicklook data base, provided through a collaboration between the Centre de Donn\'ees de la Physique des Plasmas (CDPP, supported by CNRS, CNES, Observatoire de Paris and Universit\'e Paul Sabatier, Toulouse) and Imperial College London (supported by the UK Science and Technology Facilities Council). 
\end{acknowledgements}

%-------------------------------------------------------------------
\bibliographystyle{aa}
\bibliography{refs.bib}

\begin{thebibliography}{47}
\expandafter\ifx\csname natexlab\endcsname\relax\def\natexlab#1{#1}\fi

\bibitem[{André {et~al.}(2017)André, Odelstad, Graham, Eriksson, Karlsson,
  Stenberg~Wieser, Vigren, Norgren, Johansson, Henri, Rubin, \&
  Richter}]{andre2017}
André, M., Odelstad, E., Graham, D.~B., {et~al.} 2017, Monthly Notices of the
  Royal Astronomical Society, 469, S29

\bibitem[{{Burch} {et~al.}(2007){Burch}, {Goldstein}, {Cravens}, {Gibson},
  {Lundin}, {Pollock}, {Winningham}, \& {Young}}]{burch2007}
{Burch}, J.~L., {Goldstein}, R., {Cravens}, T.~E., {et~al.} 2007, Space Sci.
  Rev., 128, 697

\bibitem[{Chamberlin {et~al.}(2008)Chamberlin, Woods, \&
  Eparvier}]{chamberlin2008}
Chamberlin, P.~C., Woods, T.~N., \& Eparvier, F.~G. 2008, Space Weather, 6, 1

\bibitem[{Deca {et~al.}(2017)Deca, Divin, Henri, Eriksson, Markidis, Olshevsky,
  \& Hor\'anyi}]{deca2017}
Deca, J., Divin, A., Henri, P., {et~al.} 2017, Phys. Rev. Lett., 118, 205101

\bibitem[{Edberg {et~al.}(2016{\natexlab{a}})Edberg, Alho, André, Andrews,
  Behar, Burch, Carr, Cupido, Engelhardt, Eriksson, Glassmeier, Goetz,
  Goldstein, Henri, Johansson, Koenders, Mandt, Möstl, Nilsson, Odelstad,
  Richter, Simon~Wedlund, Stenberg~Wieser, Szego, Vigren, \&
  Volwerk}]{edberg2016b}
Edberg, N. J.~T., Alho, M., André, M., {et~al.} 2016{\natexlab{a}}, Monthly
  Notices of the Royal Astronomical Society, 462, S45

\bibitem[{Edberg {et~al.}(2015)Edberg, Eriksson, Odelstad, Henri, Lebreton,
  Gasc, Rubin, Andr?, Gill, Johansson, Johansson, Vigren, Wahlund, Carr,
  Cupido, Glassmeier, Goldstein, Koenders, Mandt, Nemeth, Nilsson, Richter,
  Wieser, Szego, \& Volwerk}]{edberg2015}
Edberg, N. J.~T., Eriksson, A.~I., Odelstad, E., {et~al.} 2015, Geophysical
  Research Letters, 42, 2015GL064233

\bibitem[{Edberg {et~al.}(2016{\natexlab{b}})Edberg, Eriksson, Odelstad,
  Vigren, Andrews, {Johansson}, {Burch}, {Carr}, {Cupido}, {Glassmeier},
  {Goldstein}, {Halekas}, {Henri}, {Koenders}, {Mandt}, {Mokashi}, {Nemeth},
  {Nilsson}, {Ramstad}, {Richter}, \& {Wieser}}]{edberg2016}
Edberg, N.~J.~T., Eriksson, A.~I., Odelstad, E., {et~al.} 2016{\natexlab{b}},
  Journal of Geophysical Research (Space Physics), 121, 949

\bibitem[{Engelhardt {et~al.}(2018)Engelhardt, Eriksson, Vigren, Valli\'eres,
  Rubin, {Gilet, N.}, \& {Henri, P.}}]{engelhardt2018}
Engelhardt, I. A.~D., Eriksson, A.~I., Vigren, E., {et~al.} 2018, A\&A, 616,
  A51

\bibitem[{Eparvier {et~al.}(2015)Eparvier, Chamberlin, Woods, \&
  Thiemann}]{eparvier2015}
Eparvier, F.~G., Chamberlin, P.~C., Woods, T.~N., \& Thiemann, E. M.~B. 2015,
  Space Science Reviews, 195, 293

\bibitem[{{Eriksson} {et~al.}(2007){Eriksson}, {Bostr{\"o}m}, {Gill},
  {{\AA}hl{\'e}n}, {Jansson}, {Wahlund}, {Andr{\'e}}, {M{\"a}lkki}, {Holtet},
  {Lybekk}, {Pedersen}, \& {Blomberg}}]{eriksson2007}
{Eriksson}, A.~I., {Bostr{\"o}m}, R., {Gill}, R., {et~al.} 2007, Space Sci.
  Rev., 128, 729

\bibitem[{Eriksson {et~al.}(2017)Eriksson, Engelhardt, Andr\'e, Bostr\"om,
  Edberg, {Johansson, F. L.}, {Odelstad, E.}, {Vigren, E.}, {Wahlund, J.-E.},
  {Henri, P.}, {Lebreton, J.-P.}, {Miloch, W. J.}, {Paulsson, J. J. P.}, {Simon
  Wedlund, C.}, {Yang, L.}, {Karlsson, T.}, {Jarvinen, R.}, {Broiles, T.},
  {Mandt, K.}, {Carr, C. M.}, {Galand, M.}, {Nilsson, H.}, \& {Norberg,
  C.}}]{eriksson2016}
Eriksson, A.~I., Engelhardt, I. A.~D., Andr\'e, M., {et~al.} 2017, A\&A, 605,
  A15

\bibitem[{Galand {et~al.}(2016)Galand, Héritier, Odelstad, Henri, Broiles,
  Allen, Altwegg, Beth, Burch, Carr, Cupido, Eriksson, Glassmeier, Johansson,
  Lebreton, Mandt, Nilsson, Richter, Rubin, Sagnières, Schwartz, Sémon, Tzou,
  Vallières, Vigren, \& Wurz}]{galand2016}
Galand, M., Héritier, K.~L., Odelstad, E., {et~al.} 2016, Monthly Notices of
  the Royal Astronomical Society, 462, S331

\bibitem[{{Goetz} {et~al.}(2016){Goetz}, {Koenders}, {Richter}, {Altwegg},
  {Burch}, {Carr}, {Cupido}, {Eriksson}, {G{\"u}ttler}, {Henri}, {Mokashi},
  {Nemeth}, {Nilsson}, {Rubin}, {Sierks}, {Tsurutani}, {Vallat}, {Volwerk}, \&
  {Glassmeier}}]{goetz2016}
{Goetz}, C., {Koenders}, C., {Richter}, I., {et~al.} 2016, \aap, 588, A24

\bibitem[{Goetz {et~al.}(2018)Goetz, Tsurutani, Henri, Volwerk, Behar, Edberg,
  Eriksson, Goldstein, Mokashi, Nilsson, Richter, Wellbrock, , \&
  Glassmeier}]{goetz2018}
Goetz, C., Tsurutani, B.~T., Henri, P., {et~al.} 2018, Astronomy \&
  Astrophysics

\bibitem[{Grün {et~al.}(2016)Grün, Agarwal, Altobelli, Altwegg, Bentley,
  Biver, Della~Corte, Edberg, Feldman, Galand, Geiger, Götz, Grieger,
  Güttler, Henri, Hofstadter, Horanyi, Jehin, Krüger, Lee, Mannel, Morales,
  Mousis, Müller, Opitom, Rotundi, Schmied, Schmidt, Sierks, Snodgrass, Soja,
  Sommer, Srama, Tzou, Vincent, Yanamandra-Fisher, A'Hearn, Erikson, Barbieri,
  Barucci, Bertaux, Bertini, Burch, Colangeli, Cremonese, Da~Deppo, Davidsson,
  Debei, De~Cecco, Deller, Feaga, Ferrari, Fornasier, Fulle, Gicquel, Gillon,
  Green, Groussin, Gutiérrez, Hofmann, Hviid, Ip, Ivanovski, Jorda, Keller,
  Knight, Knollenberg, Koschny, Kramm, Kührt, Küppers, Lamy, Lara, Lazzarin,
  Lòpez-Moreno, Manfroid, Epifani, Marzari, Naletto, Oklay, Palumbo, Parker,
  Rickman, Rodrigo, Rodrìguez, Schindhelm, Shi, Sordini, Steffl, Stern,
  Thomas, Tubiana, Weaver, Weissman, Zakharov, \& Taylor}]{grun2016}
Grün, E., Agarwal, J., Altobelli, N., {et~al.} 2016, Monthly Notices of the
  Royal Astronomical Society, 462, S220

\bibitem[{Hajra {et~al.}(2018)Hajra, Henri, Myllys, Héritier, Galand,
  Simon Wedlund, Breuillard, Behar, Edberg, Goetz, Nilsson, Eriksson,
  Goldstein, Tsurutani, Moré, Vallières, \& Wattieaux}]{hajra2018}
Hajra, R., Henri, P., Myllys, M., {et~al.} 2018, Monthly Notices of the Royal
  Astronomical Society, 480, 4544

\bibitem[{Hajra {et~al.}(2017)Hajra, Henri, Valli\`eres, Galand, H\'eritier,
  Eriksson, Odelstad, Edberg, Burch, Broiles, Goldstein, Glassmeier, {Richter,
  I.}, {Goetz, C.}, {Tsurutani, B. T.}, {Nilsson, H.}, {Altwegg, K.}, \&
  {Rubin, M.}}]{hajra2017}
Hajra, R., Henri, P., Valli\`eres, X., {et~al.} 2017, A\&A, 607, A34

\bibitem[{Hansen {et~al.}(2016)Hansen, Altwegg, Berthelier, Bieler, Biver,
  Bockelée-Morvan, Calmonte, Capaccioni, Combi, De~Keyser, Fiethe, Fougere,
  Fuselier, Gasc, Gombosi, Huang, Le~Roy, Lee, Nilsson, Rubin, Shou, Snodgrass,
  Tenishev, Toth, Tzou, Simon~Wedlund, \& the ROSINA~team}]{hansen2016}
Hansen, K.~C., Altwegg, K., Berthelier, J.-J., {et~al.} 2016, Monthly Notices
  of the Royal Astronomical Society, 462, S491

\bibitem[{Heritier {et~al.}(2018)Heritier, Galand, Henri, Johansson, Beth,
  Eriksson, Valli\`eres, Altwegg, Burch, Carr, Ducrot, Hajra, \&
  Rubin}]{heritier2018}
Heritier, K.~L., Galand, M., Henri, P., {et~al.} 2018, A\&A, 618, A77

\bibitem[{Heritier {et~al.}(2017)Heritier, Henri, Vallières, Galand, Odelstad,
  Eriksson, Johansson, Altwegg, Behar, Beth, Broiles, Burch, Carr, Cupido,
  Nilsson, Rubin, \& Vigren}]{heritier2017}
Heritier, K.~L., Henri, P., Vallières, X., {et~al.} 2017, Monthly Notices of
  the Royal Astronomical Society, 469, S118

\bibitem[{Huang {et~al.}(2018)Huang, Tóth, Gombosi, Jia, Combi, Hansen,
  Fougere, Shou, Tenishev, Altwegg, \& Rubin}]{huang2018}
Huang, Z., Tóth, G., Gombosi, T.~I., {et~al.} 2018, Monthly Notices of the
  Royal Astronomical Society, 475, 2835

\bibitem[{Johansson {et~al.}(2017)Johansson, Odelstad, Paulsson, Harang,
  Eriksson, Mannel, Vigren, Edberg, Miloch, Simon~Wedlund, Thiemann, Eparvier,
  \& Andersson}]{johansson2017}
Johansson, F.~L., Odelstad, E., Paulsson, J. J.~P., {et~al.} 2017, Monthly
  Notices of the Royal Astronomical Society, 469, S626

\bibitem[{Judge {et~al.}(1998)Judge, McMullin, Ogawa, Hovestadt, Klecker,
  Hilchenbach, M{\"o}bius, Canfield, Vest, Watts, Tarrio, K{\"u}hne, \&
  Wurz}]{judge1998}
Judge, D.~L., McMullin, D.~R., Ogawa, H.~S., {et~al.} 1998, Solar Physics, 177,
  161

\bibitem[{Koenders {et~al.}(2015)Koenders, Glassmeier, Richter, Ranocha, \&
  Motschmann}]{koenders2015}
Koenders, C., Glassmeier, K.-H., Richter, I., Ranocha, H., \& Motschmann, U.
  2015, Planet. Space Sci., 105, 101

\bibitem[{Le {et~al.}(2012)Le, Liu, \& Wan}]{le2012}
Le, H., Liu, L., \& Wan, W. 2012, Journal of Geophysical Research: Space
  Physics, 117

\bibitem[{Lee {et~al.}(2018)Lee, Dong, Pawlowski, Thiemann, Tenishev, Mahaffy,
  Benna, Combi, Bougher, \& Eparvier}]{lee2018}
Lee, Y., Dong, C., Pawlowski, D., {et~al.} 2018, Geophysical Research Letters,
  45, 6814

\bibitem[{Mandt {et~al.}(2016)Mandt, Eriksson, Edberg, Koenders, Broiles,
  Fuselier, Henri, Nemeth, Alho, Biver, Beth, Burch, Carr, Chae, Coates,
  Cupido, Galand, Glassmeier, Goetz, Goldstein, Hansen, Haiducek, Kallio,
  Lebreton, Luspay-Kuti, Mokashi, Nilsson, Opitz, Richter, Samara, Szego, Tzou,
  Volwerk, Simon~Wedlund, \& Stenberg~Wieser}]{mandt2016}
Mandt, K.~E., Eriksson, A., Edberg, N. J.~T., {et~al.} 2016, Monthly Notices of
  the Royal Astronomical Society, 462, S9

\bibitem[{Mendillo {et~al.}(2006)Mendillo, Withers, Hinson, Rishbeth, \&
  Reinisch}]{mendillo2006}
Mendillo, M., Withers, P., Hinson, D., Rishbeth, H., \& Reinisch, B. 2006,
  Science, 311, 1135

\bibitem[{Nilsson {et~al.}(2017)Nilsson, Wieser, Behar, Gunell, Wieser, Galand,
  Simon~Wedlund, Alho, Goetz, Yamauchi, Henri, Odelstad, \&
  Vigren}]{nilsson2017}
Nilsson, H., Wieser, G.~S., Behar, E., {et~al.} 2017, Monthly Notices of the
  Royal Astronomical Society, 469, S252

\bibitem[{Noonan {et~al.}(2018)Noonan, Stern, Feldman, Broiles, Wedlund,
  Edberg, Schindhelm, Parker, Keeney, Jr, Steffl, Knight, Weaver, Feaga,
  A‚ÄôHearn, \& Bertaux}]{noonan2018}
Noonan, J.~W., Stern, S.~A., Feldman, P.~D., {et~al.} 2018, The Astronomical
  Journal, 156, 16

\bibitem[{Odelstad {et~al.}(2015)Odelstad, Eriksson, Edberg, Johansson, Vigren,
  Andr?, Tzou, Carr, \& Cupido}]{odelstad2015}
Odelstad, E., Eriksson, A.~I., Edberg, N. J.~T., {et~al.} 2015, Geophys. Res.
  Lett., 42, 10,126, 2015GL066599

\bibitem[{Peterson {et~al.}(2016)Peterson, Thiemann, Eparvier, Andersson,
  Fowler, Larson, Mitchell, Mazelle, Fontenla, Evans, Xu, Liemohn, Bougher,
  Sakai, Cravens, Elrod, Benna, Mahaffy, \& Jakosky}]{peterson2016}
Peterson, W.~K., Thiemann, E. M.~B., Eparvier, F.~G., {et~al.} 2016, Journal of
  Geophysical Research: Space Physics, 121, 8859

\bibitem[{Qian {et~al.}(2009)Qian, Burns, Chamberlin, \& Solomon}]{qian2009}
Qian, L., Burns, A.~G., Chamberlin, P.~C., \& Solomon, S.~C. 2009, Journal of
  Geophysical Research: Space Physics, 115

\bibitem[{Thiemann {et~al.}(2017{\natexlab{a}})Thiemann, Eparvier, \&
  Woods}]{thiemann2017}
Thiemann, E.~M., Eparvier, F.~G., \& Woods, T.~N. 2017{\natexlab{a}}, J. Space
  Weather Space Clim., 7, A36

\bibitem[{Thiemann {et~al.}(2018)Thiemann, Chamberlin, Eparvier, \&
  Epp}]{thiemann2018}
Thiemann, E. M.~B., Chamberlin, P.~C., Eparvier, F.~G., \& Epp, L. 2018, Solar
  Physics, 293, 19

\bibitem[{Thiemann {et~al.}(2017{\natexlab{b}})Thiemann, Chamberlin, Eparvier,
  Templeman, Woods, Bougher, \& Jakosky}]{thiemann2017b}
Thiemann, E. M.~B., Chamberlin, P.~C., Eparvier, F.~G., {et~al.}
  2017{\natexlab{b}}, Journal of Geophysical Research: Space Physics, 122, 2748

\bibitem[{Thiemann {et~al.}(2015)Thiemann, Eparvier, Andersson, Fowler,
  Peterson, Mahaffy, England, Larson, Lo, Schneider, Deighan, McClintock, \&
  Jakosky}]{thiemann2015}
Thiemann, E. M.~B., Eparvier, F.~G., Andersson, L.~A., {et~al.} 2015,
  Geophysical Research Letters, 42, 8986

\bibitem[{{Trotignon} {et~al.}(2006){Trotignon}, {Mazelle}, {Bertucci}, \&
  {Acu{\~n}a}}]{trotignon2006}
{Trotignon}, J.~G., {Mazelle}, C., {Bertucci}, C., \& {Acu{\~n}a}, M.~H. 2006,
  Planet. Space Sci., 54, 357

\bibitem[{Tsurutani {et~al.}(2009)Tsurutani, Verkhoglyadova, Mannucci, Lakhina,
  Li, \& Zank}]{tsurutani2009}
Tsurutani, B.~T., Verkhoglyadova, O.~P., Mannucci, A.~J., {et~al.} 2009, Radio
  Science, 44, RS0A17

\bibitem[{Veronig {et~al.}(2002)Veronig, Temmer, Hanslmeier, Otruba, \&
  Messerotti}]{veronig2002}
Veronig, A., Temmer, M., Hanslmeier, A., Otruba, W., \& Messerotti, M. 2002,
  A\&A, 382, 1070

\bibitem[{Vigren {et~al.}(2016)Vigren, Altwegg, Edberg, Eriksson, Galand,
  Henri, Johansson, Odelstad, Tzou, \& Vallieres}]{vigren2016b}
Vigren, E., Altwegg, K., Edberg, N. J.~T., {et~al.} 2016, The Astronomical
  Journal, 152, 59

\bibitem[{{Vigren} {et~al.}(2015){Vigren}, {Galand}, {Eriksson}, {Edberg},
  {Odelstad}, \& {Schwartz}}]{vigren2015}
{Vigren}, E., {Galand}, M., {Eriksson}, A.~I., {et~al.} 2015, Astrophys. J.,
  812:54, 9

\bibitem[{Volwerk {et~al.}(2017)Volwerk, Jones, Broiles, Burch, Carr, Coates,
  Cupido, Delva, Edberg, Eriksson, Goetz, Goldstein, Henri, Madanian, Nilsson,
  Richter, Schwingenschuh, Stenberg~Wieser, \& Glassmeier}]{volwerk2017}
Volwerk, M., Jones, G.~H., Broiles, T., {et~al.} 2017, Journal of Geophysical
  Research: Space Physics, 122, 3308

\bibitem[{Volwerk {et~al.}(2016)Volwerk, Richter, Tsurutani, G??tz, Altwegg,
  Broiles, Burch, Carr, Cupido, Delva, D??sa, Edberg, Eriksson, Henri,
  Koenders, Lebreton, Mandt, Nilsson, Opitz, Rubin, Schwingenschuh,
  Stenberg~Wieser, Szeg??, Vallat, Vallieres, \& Glassmeier}]{volwerk2016}
Volwerk, M., Richter, I., Tsurutani, B., {et~al.} 2016, Ann. Geophys., 34, 1

\bibitem[{Watanabe {et~al.}(2012)Watanabe, Masuda, \& Segawa}]{Watanabe2012}
Watanabe, K., Masuda, S., \& Segawa, T. 2012, Solar Physics, 279, 317

\bibitem[{Woods {et~al.}(2005)Woods, Eparvier, Bailey, Chamberlin, Lean,
  Rottman, Solomon, Tobiska, \& Woodraska}]{woods2005}
Woods, T.~N., Eparvier, F.~G., Bailey, S.~M., {et~al.} 2005, J. of Geophys.
  Res., 110, a01312

\bibitem[{Yang {et~al.}(2016)Yang, Paulsson, Simon~Wedlund, Odelstad, Edberg,
  Koenders, Eriksson, \& Miloch}]{yang2016}
Yang, L., Paulsson, J. J.~P., Simon~Wedlund, C., {et~al.} 2016, Monthly Notices
  of the Royal Astronomical Society, 462, S33

\end{thebibliography}

\end{document}